\documentclass[prb,twocolumn,showpacs,superscriptaddress,floatfix] {revtex4}
\usepackage{epsfig,color}
\usepackage{dcolumn}

\newcommand{\beq}{\begin{equation}}
\newcommand{\eeq}{\end{equation}}
\newcommand{\bea}{\begin{eqnarray}}
\newcommand{\eea}{\end{eqnarray}}
\newcommand{\bwt}{\begin{widetext}}
\newcommand{\ewt}{\end{widetext}}

\begin{document}

\title{ Magnetic
degeneracy and hidden metallicity of the spin density wave state in ferropnictides}

\author{I.~Eremin}
 \affiliation{Max-Planck-Institut f\"{u}r Physik komplexer Systeme, D-01187 Dresden, Germany}
 \affiliation{Institute f\"{u}r Mathematische und Theoretische Physik, TU Braunschweig, 38106 Braunschweig, Germany}
\author{A.V.~Chubukov}
 \affiliation{Department of Physics, University of Wisconsin-Madison, Madison,
Wisconsin 53706, USA}

\date{\today}

\begin{abstract}
We analyze spin density wave (SDW) order in iron-based superconductors
and electronic structure in the SDW phase.
 We consider an itinerant model for $Fe-$pnictides with
two hole bands centered at $(0,0)$ and two electron bands centered at $(0,\pi)$ and $(\pi,0)$ in  the unfolded BZ. A SDW order in such a model is generally a combination of two components with momenta $(0,\pi)$ and $(\pi,0)$,
both yield $(\pi,\pi)$ order in the folded zone.
Neutron experiments, however, indicate that only one component is present.
We show that $(0,\pi)$ or $(\pi,0)$  order is selected if we assume that only one hole band is involved in the SDW mixing with electron bands.
A SDW order in such 3-band model is highly degenerate for a perfect nesting and  hole-electron interaction only, but we show that
ellipticity of electron pockets and interactions between electron bands break the degeneracy and favor the desired $(0,\pi)$ or $(\pi,0)$ order.
We further show that stripe-ordered system remains a metal for arbitrary coupling.
We  analyze  electronic structure for parameters relevant to the pnictides and argue that the resulting electronic structure is in good agreement with ARPES experiments.
We discuss the differences between our model and $J_1-J_2$ model of localized spins.
\end{abstract}

\pacs{74.70.Xa, 75.30.Fv, 75.25.-j}

\maketitle

\section{Introduction}
\label{sec:1}

The discovery of superconductivity in the
oxypnictide  LaFeAsO (Ref. \onlinecite{kamihara}) created a
new class of  Fe-based  high-T$_c$ superconductors -- ferropnictides (FPs)
The phase diagram of FPs is similar to that of high-$T_c$ cuprates
and contains an antiferromagnetic phase at small dopings and superconducting phase at larger dopings. There are two important distinctions, however.
First, parent compounds of FPs are antiferromagnetic metals, and second,
the pairing symmetry in FPs is, most likely, an extended $s-$wave, with or without nodes. ~\cite{ext_s,maier}
Electronic structure of parent FPs in the normal state has been
measured  by angle-resolved photoemission (ARPES)~\cite{kaminski} and
by magneto-oscillations~\cite{coldea}. It
consists of two quasi-2D near-circular hole pockets of non-equal size, centered around
 $\Gamma$ point ($(0,0)$),  and two  quasi-2D elliptic electron pockets
centered around   $(0, \pm \pi)$ and $(\pm \pi,0)$  points in the
unfolded Brillouine zone (BZ) which includes only $Fe$ atoms.
 For tetragonal symmetry, the two electron pockets transform into each other under a rotation by $90^o$.
In the folded BZ, which is used for experimental measurements because of two non-equivalent
$As$ positions with respect to an  $Fe$ plane,  both electron pockets are centered around $(\pi,\pi)$.
The  dispersions near electron pockets and  near  hole pockets are
reasonably close to each other apart from the sign change, i.e., there is
a substantial degree of  nesting between hole and electron bands.
There also exists the $5$-th hole
band near $(\pi,\pi)$, but it is more three-dimensional and does not seem to play a significant role,
at least for magnetism.

In this paper we analyze spin and electronic structures of a magnetically
ordered state in parent FPs  below
$T_{SDW} \sim 150$K~\cite{kamihara,cruz,klauss}. We will only focus on
$FeAs$ materials. Neutron scattering  measurements on parent $FeAs$ pnictides
have found that the ordered momentum
in the unfolded BZ is either $(0,\pi)$ or $(\pi,0)$,
{\it i.e.},  magnetic order consists of ferromagnetic
stripes along one crystallographic direction in an $Fe$ plane,
and antiferromagnetic stripes along the other direction.

Such  magnetic order  emerges in the $J_1 - J_2$ model
of localized spins with interactions between nearest and next-nearest neighbors ($J_1$ and $J_2$, respectively),
for $J_2 > 0.5 J_1$.~\cite{chandra,si}  A localized spin model,  however,
is best suitable for an insulator and is generally not applicable to a metal
unless the system is close to a metal-insulator transition.~\cite{si}
An alternative scenario, which we explore here,  assumes that
parent $FeAs$ FPs are  ``good metals'' made of  itinerant electrons,
and antiferromagnetic order is of  spin-density-wave (SDW) type.
Given the electronic structure of FPs, it is natural to assume that
SDW order emerges, at least partly,  due to near-nesting between
the dispersions of holes and electrons~\cite{Tesanovic,lp,Chubukov2008,d_h_lee,Korshunov2008,timm,mj,honerkamp}. Such nesting is known to give rise to magnetism in $Cr$~(Ref. \onlinecite{rice}). The itinerant scenario for FPs is largely supported by a reasonable
 agreement between \textit{ab initio} electronic structure
calculations \cite{LDA} and magnetooscillation and ARPES experiments
\cite{Coldea2008,Lu2008,Liu2008,Liu2008_2,ding_latest}, although electronic structure calculations also indicate that magnetism partly comes from other fermionic bands
which do not cross the Fermi level~\cite{mazin_last}.

We  discuss below several  puzzling electronic and magnetic
features of the SDW state which have not been yet understood within the itinerant scenario. They include:\\

\begin{figure}
\includegraphics[angle=0,width=0.7\columnwidth]{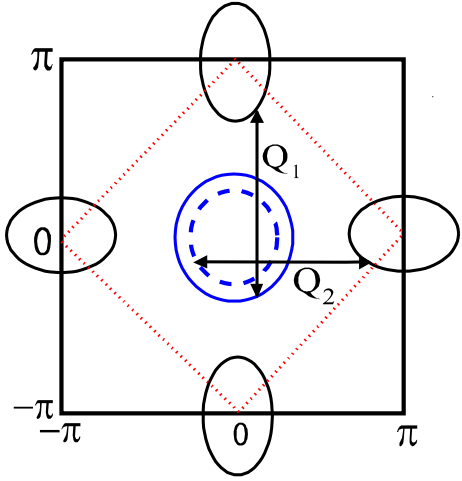}
\caption{Fermi surface
of ferropnictides in the unfolded BZ consisting of two hole pockets
centered around the $\Gamma-$point
and two electron pockets centered around $(\pi,0)$ and $(0,\pi)$ points, respectively. The
wave vectors ${\bf Q}_1$ and ${\bf Q}_2$ are two degenerate nesting wave vectors.} \label{fig1}
\end{figure}

\begin{itemize}
\item
A higher conductivity in the SDW state --
the system becomes more metallic below $T_{SDW}$. Naively, one would expect
a smaller conductivity due to at least partial gapping of the Fermi surfaces (FS). A reduction of the quasiparticle damping, induced by such gapping,
 may slow down the decrease of conductivity,
but the observed increase is highly  unlikely  for a conventional SDW scenario.
\item
A significant reconstruction of the electronic dispersion across the SDW transition and complicated FS topology below T$_{SDW}$, with more FS crossings than in the normal state (Refs. \cite{kondo,zhou,dresden_arpes,Yi}) Some ARPES measurements show\cite{kondo,zhou,dresden_arpes}
that in the SDW state visible dispersion remains hole-like around the $\Gamma-$point despite apparent mixing of electron and hole bands,
while close to  $M=(\pi,\pi)$  both electron and hole bands are present.
Other measurements~\cite{Yi} indicate that hole and  electron pockets are present both near $M$ and near $\Gamma$ points.
\item
The particular $(0,\pi)$ ($(\pi,0)$)
ordering of Fe spins in the unfolded BZ. It is
straightforward to obtain $(\pi,\pi)$ ordering in the folded BZ simply because
hole and electron bands are shifted by $(\pi,\pi)$
in the folded zone. However, SDW order in the folded zone
involves two separate magnetic sublattices,
and  unfolding reveals two SDW order parameters (OP)
$\vec{\Delta}_1$ and $\vec{\Delta}_2$ with momenta ${\bf Q}_1 = (0,\pi)$ and
${\bf Q}_2 =(\pi,0)$, respectively (see Fig. \ref{fig1})
A generic
spin configuration  then has the form $ {\vec S} ({\bf R}) = {\vec\Delta}_1 e^{i{\bf Q}_1 {\bf R}}+{\vec \Delta}_2 e^{i{\bf Q}_2 {\bf R}}$ (the sublattice OPs are $\vec{\Delta}_1 + \vec{\Delta}_2$ and
$\vec{\Delta}_1 - \vec{\Delta}_2$).
For such configuration, next-nearest neighbors are antiferromagnetically oriented, but the length and the direction of ${\vec S}$ between nearest neighbors varies.
Only if either ${\vec \Delta}_1$ or ${\vec \Delta}_2$ vanish, OPs of the
two sublattices align parallel or antiparallel to each other, and the
spin configuration becomes the same as in the experiments. The issue then is what interaction causes ${\vec \Delta}_1$ or ${\vec \Delta}_2$ to vanish.
\end{itemize}

Another unsolved issues which we do not address here
are the absence of magnetism in the nominally undoped
LaFePO and LiFeAs compounds and the type of a magnetic order in Fe$_{1+x}$Te/Se systems.
The absence of magnetism in  LaFePO and LiFeAs  could be due to the fact that these materials are less quasi-two-dimensional, with less nesting. \cite{coldea_nesting}  For Fe$_{1+x}$Te/Se, the experimental data are still controversial. \cite{fetese}

The full analysis of possible SDW orderings
in $FeAs$ systems within a generic 4-band model is quite
messy, and one should have a good staring point to be able to understand the
physics.  One way to analyze the problem is to solve first the
model of 4 equivalent 2D
circular bands (two hole bands and two electron bands) with isotropic
interband and intraband interactions, and then introduce
anisotropy between the two hole bands,  ellipticity of the two electron bands, and  anisotropy of the interactions.
This approach has been put forward by Cvetkovic and
Tesanovic~\cite{CT09}. They solved exactly the isotropic model and
found that SDW state is an insulator -- all
4 Fermi surfaces are  fully gapped,  and that there exists a degenerate manifold of SDW orders. The degeneracy is the same as in $J_2$ model of localized spins: the SDW order involves two antiferromagnetic sublattices with equal magnitudes of the OPs (i.e., ${\vec \Delta}_1 \cdot {\vec \Delta}_2 =0$), but the angle between the two sublattices $\phi = \cos^{-1} (\Delta^2_1 - \Delta^2_2)/(\Delta^2_1 + \Delta^2_2)$ can be arbitrary.   The $(0,\pi)$ and $(\pi,0)$ states belong to this manifold, but are not yet selected in the isotropic model.
The issue not yet addressed within this approach is what happens with a degenerate manifold once one moves away from the isotropic limit.

Our starting point is different. We use the experimental fact
that the two  hole FSs are of quite different sizes\cite{zhou} and assume that
one hole band interacts with electron bands much
stronger than the other. We then consider, as a first approximation, a
3-band model of one hole band centered around $\Gamma$-point and two
electron bands centered around ($0,\pi$) and $(\pi,0)$.  We solve this model in the
mean-field approximation and show that the SDW order is still degenerate
for isotropic interactions and circular bands.
A degenerate manifold consists of two antiferromagnetic sublattices with generally non-equal magnitudes of the
sublattice OPs. The manifold
includes $(0,\pi)$ and $(\pi,0)$ states among many others.

In terms of ${\vec \Delta}_1$ and ${\vec \Delta}_2$, the degeneracy implies that the ground state energy depends only on the combination
$(\vec{\Delta}_1)^2 + (\vec{\Delta}_2)^2$. The excitation spectrum of such degenerate state  contains five Goldstone modes. We then show that
the  ellipticity of the electron bands and the anisotropy of the interactions
breaks the degeneracy  and adds to the energy the
term $\beta_{12} |\vec{\Delta}_1|^2 |\vec{\Delta}_2|^2$ with positive
$\beta_{12}$. The full energy is then minimized when either $\vec{\Delta}_1 =0$, or $\vec{\Delta}_2 =0$, {\it  ellipticity and the anisotropy of the interactions both select
 $(0,\pi)$  or $(\pi,0)$  stripe order already at a mean-field level}.
This is one of the main results of this paper.
A stripe SDW order mixes the hole band and one of the two electron bands, but leaves another electron band intact. As a result, the  system remains a metal even at strong SDW coupling.
We verified that the
selection of the stripe order by the interactions between electron states
holds only if the interactions are in the charge channel.
The same interactions, but in the spin channel,
select a different state with ${\vec \Delta}_1 = \pm {\vec \Delta}_2$, in which SDW order resides in only one of the two  sublattices, see Fig. \ref{fig2}(d).

We did not consider
corrections to the mean-field theory (i.e., quantum fluctuations). They are
relatively small, at least at weak coupling, although,  very likely,
they also break $O(6)$ symmetry.
We also do not consider the coupling to the phonons, which is
another potential source of symmetry breaking and also pre-emptive tetragonal to orthorhombic transition.\cite{barzykin}.

We next consider the role of the second hole band. This band
interacts with the electron band left out of the primary SDW mixing.
These two bands  are less nested,
and the interaction must exceed a threshold for
an additional SDW order to appear. If this is the case and an
additional SDW order is strong enough,
it gaps  the remaining two FSs, and the system  becomes an insulator.
This second (weaker) SDW OP has the momentum $(0,\pi)$ if the original SDW order was with $(\pi,0)$, and vise versa. It then introduces ${\vec \Delta}_2$ (or ${\vec \Delta}_1$) which was set to zero by the initial selection of $(0,\pi)$ or $(\pi,0)$ order.
We show that this second order parameter is directed
orthogonal to the primary one, {\it i.e.}, ${\vec \Delta}_1 \cdot {\vec \Delta}_2 =0$.
The magnitude of the OP at each $Fe$ cite is then the same,
but the OP is not
ferromagnetic  along $x$ or $y$  direction, {\it i.e.}, the resulting SDW order is different from $(0,\pi)$ or$(\pi,0)$.
The only way to preserve $(0,\pi)$ or $(\pi,0)$  order in the 4-band model is
to assume that the interaction between the second hole FS and the electron FS left out in primary SDW selection is below the threshold. Then  ${\vec \Delta}_2$ (or ${\vec \Delta}_1$) remains zero,
and $(0,\pi)$ or $(\pi,0)$ order survives.  A simple but important consequence
of this observation is that the 4-band model with a stripe order must remain a metal. Specifically,
 one hole and one electron band
are not involved in the SDW mixing and  should be observed in  ARPES experiments exactly at the same positions as in the normal state: the hole band near $(0,0)$ and the electron band near $(\pi,\pi)$. This somewhat unexpected result
is another main conclusion of this paper.

Applying the results to the pnictides,
where the interactions are believed to be moderate,
we show that the hole band with a larger FS interacts more strongly with
elliptic electronic bands than the hole band with a smaller FS.
Hence, in our theory, the smaller hole FS stays intact below a SDW transition, and
the reconstruction of the fermionic structure below $T_{SDW}$
involves the larger hole FS and one of two electron FSs. Since  the hole FS is a circle
and the electron FS are ellipses,  the two cross at four points when shifted by, {\it e.g.},
$(0,\pi)$, and for moderately-strong  interactions SDW  gaps
open up only around crossing points~\cite{parker}.
In the folded BZ this gives rise to two FS crossings near either $(0,0)$ or $(\pi,\pi)$. Adding the FSs which are not involved in the SDW mixing, we find that there generally must be 3 FS crossings near $(0,0)$ or $(\pi,\pi)$ -- more than in the normal state.

Further, our conclusion that  $(0,\pi)$ SDW order in the 4-band model preserves  metallic behavior even at strong coupling implies that there is no continuous evolution between our 4-band model
and $J_1-J_2$ model of localized spins, despite that quantum fluctuations in $J_1-J-2$ model select the same $(0,\pi)$ order. At a first glance, this is surprising as it is well-known that in the one-band Hubbard model, there is an evolution from itinerant to localized behavior~\cite{swz} by which we mean
that, upon increasing $U$, a one-band system evolves
from  an antiferromagnetic metal to an antiferromagnetic insulator with the same with $(\pi,\pi)$ magnetic order. To understand why our 4-band model is special, we considered a half-filled $t-t'-U$
Hubbard model on a square lattice with the hopping $t$ between nearest neighbors  and hopping $t'$ between next-nearest neighbors along the diagonals.  In the large $U$ limit, the model reduces to $J_1-J_2$ model with $J_1 \sim t^2/U$ and $J_2 \sim (t')^2/U$. As small $U$, the model  describes two hole bands centered at $(0,0)$ and $(\pi,\pi)$,  and two electron bands centered at $(\pi,0)$ and $(0,\pi)$.  At $t=0$, all four bands are identical (up to an overall
sign), and all  have an isotropic quadratic dispersion near the top of the hole bands and the bottom of the
electron bands. For $t \neq 0$, electron bands become elliptical near the bottom, and the masses of the two hole bands become different ({\it i.e.}, one hole FS becomes larger and the other becomes smaller).  This very much resembles the
geometry of our 4-band model, except that in $t-t'-U$ model
the second hole band is located at $(\pi,\pi)$ (see Fig.5). We analyzed $t-t'-U$ model in the same way as the 4-band model and found that in $t-t'-U$ model there is an evolution from a metallic to an insulating behavior as $U$ increases. Namely, the state with $(0,\pi)$ or $(\pi,0)$  stripe order is a metal at small $U$ and an
insulator at large $U$.

The difference between our case and $t-t'-U$ model can be easily understood.
If we apply the same logics as in our case, {\it i.e.},
consider the 3-band model first, we find $(0,\pi)$ SDW order which
mixes one hole and one electron FSs.
The remaining electron band can again mix with the other hole band
and create another SDW order, what eventually gaps all 4 FSs. However, in $t-t'-U$ model, the second hole FS is located at $(\pi,\pi)$ rather than at the $\Gamma$ point, and the second SDW OP has the same momentum $(0,\pi)$ as
the primary SDW OP. As a result, the two OPs just add up, and $(0,\pi)$ order survives.  From this perspective,  the reason why $(0,\pi)$ ordered state in the pnictides  is a metal is the FS geometry:
the fact that both hole
FSs are centered at the same $\Gamma$ point. Were they centered at $(0,0)$ and $(\pi,\pi)$ the system would be a $(0,\pi)$ metal at weak coupling and $(0,\pi)$ insulator, described by $J_1-J_2$ model, at strong coupling.

In the rest of the paper we discuss these issues in full detail.
The structure of the presentation is the following. In Sec. \ref{sec:2_0}
we introduce three-band model and discuss SDW ordering in general terms. In
Sec. \ref{sec:2a} we consider the case of circular electron pockets and only the interactions between hole and electron pockets.
We show that, at this level, there exists a degenerate manifold of SDW ordered states. In Sec. \ref{sec:3} we show that ellipticity of electron pockets and interactions between them
remove the degeneracy and select a stripe SDW order already at the mean-field level. In Sec.
\ref{subsubsec:4} we consider spin component of the interaction between electron pockets and show that it selects a different state in which SDW order appears only at a half of Fe sites.

In Sec. \ref{sec:3a} we add the second hole
band and discuss a potential SDW mixing between this
band and the electron band left out of SDW mixing in the 3 band model.
We argue that $(0,\pi)$ or $(\pi,0)$ order survives only if this
this additional SDW mixing
does not happen, {\it i.e.}, the interaction between these two bands is below the threshold.
In Sec. \ref{sec:t-t'} we contrast this behavior with the one in a $t-t'-U$ Hubbard model with nearest and next-nearest neighbors. We show that in this model the second hole FS is centered at $(\pi,\pi)$ rather than at $(0,0)$, and
the secondary SDW OP has the same momentum as the primary OP, i.e., the
stripe order is preserved.
We show that $(0,\pi)$ or $(\pi,0)$ state in  $t-t'-U$ model
can be either a metal or an insulator, described at strong coupling
by $J_1-J_2$ model.  In Sec. \ref{sec:5}
we consider electronic structure of the SDW phase with $(0,\pi)$ order and compare our theoretical results with the experiments. We argue that the agreement with ARPES is quite good.  We present our conclusions in Sec. \ref{sec:6}.

Our results are complimentary to earlier arguments as to why
SDW phase remains a metal.  Ran {\it et al.}\cite{vishwanath}
considered an orbital model
and argued that the SDW gap must at least have nodes because
SDW coupling  vanishes along particular directions in $k-$space.
Cvetkovic and Tesanovic~\cite{Tesanovic} and Vorontsov {\it et al.} \cite{vorontsov} argued that a metal survives even for the case of circular FSs,
if SDW order is incommensurate.  As we said, we argue that
the system described by 4-band model remains a metal
even if the order is commensurate $(0,\pi)$ or $(\pi,0)$, and the interactions are angle-independent. Our analysis also serves as a justification to the works\cite{brydon,tohyama}
which discussed the  consequences of the $(\pi,0)$ SDW order in
itinerant models without analyzing why this order is selected.
Some of our results are also consistent with the
analysis of possible magnetically ordered states in the two-orbital model by Lorenzana {\it et al}.\cite{jose}. We also found, for a particular model, a state
with magnetic order residing only on one of the two sublattices. Such state was identified as a possible candidate in a generic Landau-theory analysis for a two-sublattice order paramter.~\cite{jose}

\section{Itinerant SDW order in the 3 band model}

\subsection{General consideration}
\label{sec:2_0}
Consider a model of
interacting fermions with a circular hole FS
centered around  $\Gamma$-point ($\alpha$-band)
and two elliptical electron Fermi surface pockets centered around $(\pm \pi, 0)$ and $(0, \pm \pi)$ points in the unfolded BZ ($\beta$-bands) (see Fig.\ref{fig1}):
\begin{eqnarray}
\label{eqH}
\lefteqn{H_2  =} && \nonumber\\
&& \sum_{\mathbf{p}, \sigma} \left[ \varepsilon^{\alpha_1}_{\mathbf{p}} \alpha_{i\mathbf{p}  \sigma}^\dag \alpha_{i\mathbf{p} \sigma}
+ \varepsilon^{\beta_{1}}_{\mathbf{p}} \beta_{1\mathbf{p} \sigma}^\dag \beta_{1\mathbf{p} \sigma} + \varepsilon^{\beta_{2}}_{\mathbf{p}} \beta_{2 \mathbf{p} \sigma}^\dag \beta_{2\mathbf{p} \sigma} \right]. \nonumber\\
\end{eqnarray}
Here, $\varepsilon^{\alpha_{1}}_{\mathbf{p}} = - \frac{\hbar^2 p^2}{2 m_1}
+\mu $ and  $\varepsilon^{\beta_1}_{\mathbf{p}}= \frac{\hbar^2p_x^2 }{2 m_{x}} + \frac{\hbar^2 p_y^2 }{2 m_{y}} -\mu $, $\varepsilon^{\beta_2}_{\mathbf{p}}= \frac{\hbar^2 p_x^2}{2 m_{y}} +\frac{\hbar^2 p_y^2}{2 m_{x}} -\mu $ are the dispersions of hole and electron bands.
The momenta of $\alpha-$ fermions are counted from $(0,0)$, the momenta of
$\beta_1-$ and $\beta_2-$fermions are counted  from $(0,\pi)$ and ($\pi,0$), respectively.

The interacting part of the Hamiltonian contains
density-density interactions involving $\alpha$ and $\beta$ fermions
(interactions with small momentum transfer),
 $(\pi,0)$, $(0,\pi)$, and $(\pi,\pi)$ scattering processes,
and umklapp pair hopping  terms \cite{Chubukov2008}.
 Consider first only the interactions between hole
 and electron bands,
which give rise to a SDW order.
These are density-density interaction between $\alpha$ and $\beta$ fermions, and the pair hopping term~\cite{Chubukov2008}:
\begin{eqnarray}
&&H_{4}=U_1 \sum  {\alpha}^{\dagger}_{1{\bf p}_3 \sigma}
{\beta}^{\dagger}_{j{\bf p}_4 \sigma'}  {\beta}_{j{\bf p}_2 \sigma'} {\alpha}_{1{\bf p}_1
\sigma}  + \nonumber \\
&& \frac{U_3}{2}~ \sum
\left[{\beta}^{\dagger}_{j{\bf p}_3 \sigma} {\beta}^{\dagger}_{j{\bf p}_4 \sigma'}
{\alpha}_{1{\bf p}_2 \sigma'} {\alpha}_{1{\bf p}_1 \sigma} + h.c \right],
 \label{eq:2}
\end{eqnarray}
 We neglect potential angular dependencies of $U_1$ and $U_3$ along the FSs.

Because  $U_1$ and $U_3$ involve
$\beta_1$ and $\beta_2$ fermions, we have to introduce two SDW OPs
${\vec \Delta}_1 \propto \sum_{\bf p} \langle \alpha^\dag_{1\mathbf{p}\delta} \beta_{1\mathbf{p} \gamma}  \vec{\sigma}_{\delta \gamma} \rangle$  with momentum  ${\bf Q}_1 = (0,\pi)$ and ${\vec \Delta}_2 \propto \sum_{\bf p} \langle \alpha^\dag_{1\mathbf{p}\delta} \beta_{2\mathbf{p} \gamma}  \vec{\sigma}_{\delta \gamma} \rangle$
with momentum  ${\bf Q}_2 = (\pi,0)$. Without loss of generality we can
set $\vec{\Delta}_1$ along $z$-axis and
$\vec{\Delta}_2$ in the $xz$-plane. In explicit form, we introduce
\bea
&&\Delta_{1}^{z}= -
U_{SDW}
 \sum_{\bf p} \langle \alpha^\dag_{1\mathbf{p}\uparrow} \beta_{1\mathbf{p} \uparrow}  \rangle \nonumber \\
&&\Delta_{2}^{z(x)}= -
 U_{SDW} \sum_{\bf p} \langle \alpha^\dag_{1\mathbf{p}\uparrow} \beta_{2\mathbf{p}\uparrow(\downarrow)}  \rangle.
\label{f_1_1}
\eea
where $U_{SDW} = U_1 + U_3$.

\subsection{Degeneracy of the SDW order}
\label{sec:2a}

\begin{figure}
\includegraphics[angle=0,width=0.8\columnwidth]{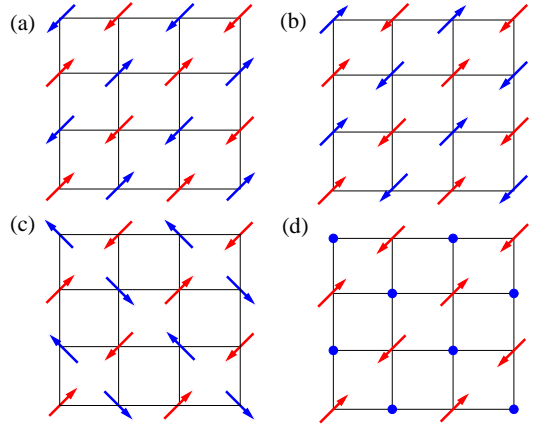}
\caption{(color online) Various SDW spin configurations
  described by $\vec{\Delta}_1 e^{i{\bf Q}_1 R}+\vec{\Delta}_2 e^{i{\bf Q}_2 R}$.
For the model of Eq. (\protect\ref{eq:2}), only  $\vec{\Delta}^2_1 +
\vec{\Delta}^2_2$ is fixed. Panel (a) --  $\vec{\Delta}_1 = 0$,
panel (b) --  $\vec{\Delta}_2 = 0$, panel (c) --   $\vec{\Delta}_1 \perp \vec{\Delta}_2$, and panel (d) -- $\vec{\Delta}_1 = \vec{\Delta}_2$} \label{fig2}
\end{figure}

Assume first that electron pockets are circular, {\it i.e.}, $m_x = m_y$ and
$\varepsilon^{\beta_{1}}_{\mathbf{p}} = \varepsilon^{\beta_{2}}_{\mathbf{p}} = \varepsilon^{\beta}_{\mathbf{p}}$.
In this situation, the Hamiltonian, Eq. (\ref{eq:2}), can
be easily diagonalized by performing three subsequent Bogolyubov
transformations. First
\begin{eqnarray}
\beta_{2{\bf p}\uparrow}=\beta_{2{\bf p}a} \cos \eta - \beta_{2{\bf p}b} \sin \eta \nonumber\\
\beta_{2{\bf p}\downarrow}=\beta_{2{\bf p}a} \sin \eta + \beta_{2{\bf p}b} \cos \eta ,
\end{eqnarray}
with $\cos \eta = \frac{\Delta_2^z}{\Delta_2}$, $\sin \eta = \frac{\Delta_2^x}{\Delta_2}$ and $\Delta_2=\sqrt{\left(\Delta_2^x\right)^2+\left(\Delta_2^z\right)^2}$. Then
\begin{eqnarray}
\beta_{1{\bf p}\uparrow}=c_{a{\bf p}} \cos \theta - d_{a{\bf p}} \sin \theta, \nonumber\\
\beta_{2{\bf p}a}=c_{a{\bf p}} \sin \theta + d_{a{\bf p}} \cos \theta, \nonumber\\
\beta_{1{\bf p}\downarrow}=c_{b{\bf p}} \cos \theta - d_{b{\bf p}} \sin \theta, \nonumber\\
\beta_{2{\bf p}b}=c_{b{\bf p}} \sin \theta + d_{b{\bf p}} \cos \theta,
\end{eqnarray}
with $\cos \theta = \frac{\Delta_1^z}{\Delta}$, $\sin \theta = \frac{\Delta_2}{\Delta}$ and $\Delta=\sqrt{\left(\Delta_2\right)^2+\left(\Delta_1^z\right)^2}$.
And, finally,
\begin{eqnarray}
\alpha_{1{\bf p}\uparrow} & = & p_{a {\bf p}} \cos \psi + e_{a {\bf p}} \sin \psi,\nonumber\\
\alpha_{1{\bf p}\downarrow}& = & p_{b {\bf p}} \cos \tilde{\psi} + e_{p{\bf p}} \sin \tilde{\psi},
\label{f_5}
\end{eqnarray}
and
\begin{eqnarray}
c_{a{\bf p}} & = & e_{a {\bf p}} \cos \psi - p_{a {\bf p}} \sin \psi, \nonumber\\
c_{b{\bf p}}& = & e_{b {\bf p}} \cos \tilde{\psi} - p_{b {\bf p}} \sin \tilde{\psi},
\label{f_6}
\end{eqnarray}
where $\tilde{\psi}=\frac{\pi}{2} +\psi$,  $\cos^2 \psi, ~\sin^2 \psi =
\frac{1}{2} \left[ 1 \pm \frac{\varepsilon_{\bf
p}}{\sqrt{\left(\varepsilon_{\bf p}\right)^{2}+\Delta^{2}}} \right]$, and
 $\varepsilon_{\bf p}=\frac{1}{2} \left[\varepsilon^{\alpha_{1}}_{\bf p} - \varepsilon^\beta_{\bf p}\right]$
The resulting quadratic Hamiltonian becomes
\bea
\lefteqn{H^{eff}_2 =  \sum_{a,{\bf p}}  \varepsilon^{\beta}_{{\bf p}}  d^\dagger_{a{\bf p}}  d_{a{\bf p}} +}
&& \nonumber \\
&& \sum_{p} E_{{\bf p}} \left( e^\dagger_{a{\bf p}}  e_{a{\bf p}} + p^{\dagger}_{b{\bf p}}  p_{b{\bf p}} - e^\dagger_{b{\bf p}}  e_{b{\bf p}} - p^{\dagger}_{a{\bf p}}  p_{a{\bf p}}\right),
\label{eq:n7}
\eea
where
$E_{\bf p} =  \pm \sqrt{\left(\varepsilon_{\bf p}\right)^2+|\Delta|^2}$.
The self-consistent equation for the gap reduces to
\beq
1 = \frac{
U_{SDW}}{2N} \sum_{\bf p} \frac{1}{\sqrt{\left(\varepsilon^-_{\bf p}\right)^{2}+\Delta^{2}}}.
\label{eq:n2}
\eeq
Two key observations follow from these results. First,
self-consistency equation Eq.(\ref{eq:n2}) sets the value of the total order parameter ${\vec \Delta}^2_1 + {\vec \Delta}^2_2$ but does not specify
what ${\vec \Delta}_1$ and ${\vec \Delta}_2$ are. The implication is that,
at this level, the OP manifold can be viewed as a $6-$ component vector (three components of
${\vec \Delta}_1$ and 3 of ${\vec \Delta}_2$), SDW ordering is a spontaneous breaking of
$O(6)$ symmetry, and an ordered state has five Goldstone modes. The degenerate
ground state manifold is composed of two-sublattice states
with antiferromagnetic order along diagonals; $(0,\pi)$ or $(\pi,0)$ states, for which ${\vec \Delta}_1 =0$ or ${\vec \Delta}_2 =0$, are just two of many possibilities (see Fig. \ref{fig2}).
Second, a linear combination of original electronic operators described by $d_{a,{\bf p}}$ decouples from the SDW procedure. Then, even in case of perfect nesting, when $\varepsilon^{\alpha_1}_{\bf p} = -
\varepsilon^\beta_{{\bf p}}$, the system still remains a metal in the SDW phase -- excitations described by $e_{a,{\bf p}}$ and $p_{a,\bf p}$ operators become gapped, but excitations
described by $d_{a,{\bf p}}$ operators remain gapless.

Note that the ground state degeneracy is even larger than in the $J_1-J_2$ model of localized spins -- not only the angle between the two sublattices can be arbitrary but also
the magnitudes of the ordered moments in the two sublattices can be different.
To confirm this result,  we computed the ground state energy  $E_{gr}$  in the mean-field approximation and indeed found that it depends only on ${\vec \Delta}^2_1 + {\vec \Delta}^2_2$.
We also extended this analysis to include $\alpha-\beta$ interactions with momentum transfer $(0,\pi)$ and $(\pi,0)$ ($U_2$ terms in the terminology of Ref.
\cite{Chubukov2008})  and still found the same degeneracy.

Note by passing that the momentum integration in the self-consistency condition, Eq. (\ref{eq:n2})  is {\it not} restricted to the FS, and a finite
SDW order parameter $\Delta$  appears
even when FS disappear. This is due to the fact that a particle-hole bubble made out of $\alpha$ and $\beta$ fermions actually behaves as a particle-particle bubble because of
sign difference  of fermionic dispersions of $\alpha$ and $\beta$ fermions.
The consequence is that the SDW ordered moment does not scale with the (small) sizes of the hole and electrons FSs.
In other words,  SDW order is due to dispersion nesting of $\alpha$ and $\beta$ fermions, but {\it not due to  FS nesting}.
\begin{figure}
\includegraphics[angle=0,width=1.0\columnwidth]{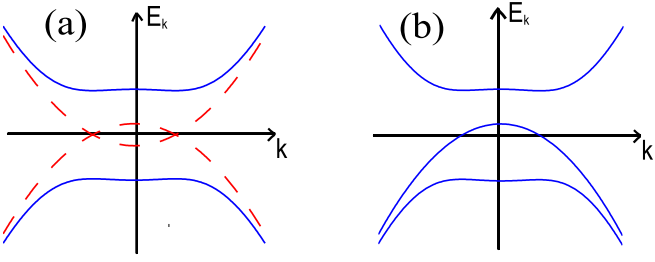}
\caption{(color online) Energy dispersions in the SDW state for full nesting:
(a) -- one electron band and one hole band, (b) -- one electron band and two hole bands.
Dashed lines in (a) are the dispersions without SDW order.} \label{fig3}
\end{figure}

\subsection{Selection of the SDW order}
\label{sec:3}

\subsubsection{Interaction between electron pockets}

We now add  the interactions between the two electron pockets
and verify whether they lift the degeneracy.
These interaction do not contribute to the quadratic form, but they do contribute to the $\Delta$-dependence of  the ground state energy. It has been argued~\cite{maier},
based on the transformation of the underlying orbital model into a band model, that the interactions between  the two electron pockets  are not particularly small and must be included into the theory.

Thee are four possible $\beta-\beta$ interactions:
\begin{widetext}
\bea
&&H^{ex}_4 = U_{6} \sum {\beta}^{\dagger}_{1{\bf p}_3 \sigma}
{\beta}^{\dagger}_{2{\bf p}_4 \sigma'}  {\beta}_{2{\bf p}_2 \sigma'} {\beta}_{1{\bf p}_1
\sigma} + U_{7} \sum {\beta}^{\dagger}_{2{\bf p}_3 \sigma}
{\beta}^{\dagger}_{1{\bf p}_4 \sigma'} {\beta}_{2{\bf p}_2 \sigma'} {\beta}_{1{\bf p}_1
\sigma} \nonumber \\
&& + \frac{U_{8}}{2} \sum
\left[{\beta}^{\dagger}_{2{\bf p}_3 \sigma} {\beta}^{\dagger}_{2{\bf p}_4 \sigma'}
{\beta}_{1{\bf p}_2 \sigma'} {\beta}_{1{\bf p}_1 \sigma} + h.c \right] +  \frac{U_{4}}{2} \sum \left[ \sum  {\beta}^{\dagger}_{1{\bf p}_3 \sigma}
{\beta}^{\dagger}_{1{\bf p}_4 \sigma'} {\beta}_{1{\bf p}_2 \sigma'} {\beta}_{1{\bf p}_1\sigma} +
  {\beta}^{\dagger}_{2{\bf p}_3 \sigma}
{\beta}^{\dagger}_{2{\bf p}_4 \sigma'} {\beta}_{2{\bf p}_2 \sigma'} {\beta}_{2{\bf p}_1\sigma}  \right]
\label{eq:n4}
\eea
\end{widetext}
(we used the terminology consistent with Ref. \cite{Chubukov2008}).
It is natural to assume that all interactions are repulsive, {\it i.e.}, all $U_i >0$.

Applying the sequence of Bogolyubov transformations and taking $\langle \cdots \rangle $, we obtain the contribution to the ground state energy from
various terms in Eq. (\ref{eq:n4}):
\bwt
\bea
&& U_{6} \sum {\beta}^{\dagger}_{1{\bf p}_3 \sigma}
{\beta}^{\dagger}_{2{\bf p}_4 \sigma'}  {\beta}_{2{\bf p}_2 \sigma'} {\beta}_{1{\bf p}_1
\sigma} \rightarrow  2 A^2  \frac{|\vec{\Delta}_1|^2 |\vec{\Delta}_2|^2}{\Delta^4}  + ...\nonumber \\
&& U_{7} \sum {\beta}^{\dagger}_{2{\bf p}_3 \sigma}
{\beta}^{\dagger}_{1{\bf p}_4 \sigma'} {\beta}_{2{\bf p}_2 \sigma'} {\beta}_{1{\bf p}_1
\sigma} \rightarrow   2 A^2 \left(2 \frac{\left(\vec{\Delta}_1 \cdot \vec{\Delta}_2 \right)^2} {\Delta^4} -  \frac{|\vec{\Delta}_1|^2 |\vec{\Delta}_2|^2}{\Delta^4}\right) + ... \nonumber \\
&& U_{8} \sum
\left[{\beta}^{\dagger}_{2{\bf p}_3 \sigma} {\beta}^{\dagger}_{2{\bf p}_4 \sigma'}
{\beta}_{1{\bf p}_2 \sigma'} {\beta}_{1{\bf p}_1 \sigma} + h.c \right] \rightarrow
4 A^2  \frac{|\vec{\Delta}_1|^2 |\vec{\Delta}_2|^2}{\Delta^4} \nonumber \\
&&U_{4} \sum \left[ \sum  {\beta}^{\dagger}_{1{\bf p}_3 \sigma}
{\beta}^{\dagger}_{1{\bf p}_4 \sigma'} {\beta}_{1{\bf p}_2 \sigma'} {\beta}_{1{\bf p}_1\sigma} +
  {\beta}^{\dagger}_{2{\bf p}_3 \sigma}
{\beta}^{\dagger}_{2{\bf p}_4 \sigma'} {\beta}_{2{\bf p}_2 \sigma'} {\beta}_{2{\bf p}_1\sigma} \right] \rightarrow -4 A^2  \frac{|\vec{\Delta}_1|^2 |\vec{\Delta}_2|^2}{\Delta^4}
\label{su_8}
\eea
\ewt
where dots stand for the terms which depend only on $\Delta^2$ and do not break a degeneracy, and  $A = (c-d)$, where $c = \langle c^{\dag}_{b{\bf p}} c_{b{\bf p}} \rangle = \langle c^{\dag}_{a{\bf p}} c_{a{\bf p}} \rangle = \frac{1}{2N}\sum_{\bf p} \left(1-\frac{\varepsilon^-_{\bf
k}}{\sqrt{\left(\varepsilon_{\bf k}\right)^{2}+\Delta^{2}}}\right)$ and $d\equiv \langle d^{\dag}_{b{\bf p}} d_{b{\bf p}} \rangle = \langle d^{\dag}_{a{\bf p}} d_{a{\bf p}} \rangle = \frac{1}{2N}\sum_{\bf p} \left(1-\frac{E_{\bf
k}}{|E_{\bf k}|}\right)$.
The quantity $A$ is zero in the normal state, but has a finite value in the SDW state.
Combining all contributions, we obtain
\begin{eqnarray}
E^{ex}_{gr} & = & 2 A^2 \left[\left(U_{6} + U_8 - U_{7} - U_{4}\right)\right] \frac{|\vec{\Delta}_1|^2 |\vec{\Delta}_2|^2}{\Delta^4} + \nonumber \\
&& 4 A^2 U_{7} \frac{\left(\vec{\Delta}_1 \cdot \vec{\Delta}_2 \right)^2} {\Delta^4}
\label{energy}
\end{eqnarray}
We see that $E^{ex}_{gr}$ depends on $|\vec{\Delta}_1|^2 |\vec{\Delta}_2|^2$ and
on  $\left(\vec{\Delta}_1 \cdot \vec{\Delta}_2 \right)^2$, {\it i.e.},
it is sensitive to both,  relative values and relative
directions of ${\vec \Delta}_1$ and ${\vec \Delta}_2$. When
 all interactions are of equal strength, the first term vanishes, and the
last term favors ${\vec \Delta}_1 \perp {\vec \Delta}_2$.
In this situation,  the $O(6)$ degeneracy is broken, but
only down to $O(3) \times O(2)$, {\it i.e.},
the magnitude of the order parameter at each site  is now the same because
 $\left({\vec \Delta}_1+{\vec\Delta}_2\right)^2=\left({\vec \Delta}_1-{\vec \Delta}_2\right)^2$, but
the angle between the directions of the SDW order in the two sublattices
(i.e., between ${\vec\Delta}_1 +{\vec\Delta}_2$ and $\vec{\Delta}_1
-\vec{\Delta}_2$) is still arbitrary. This is exactly the same situation as
in the classical $J_1-J_2$ model.  However, once $U_6+ U_8-U_7-U_4$ is nonzero, the degeneracy is broken down to a conventional $O(3)$ already at the mean-field level.
Because $U_4$ is reduced and even changes sign under RG
~\cite{Chubukov2008}, while other $U_i$ do not flow, the most likely situation is that  $U_6+ U_8-U_7-U_4 >0$, in which case $E^{ex}_{gr}$ is minimized when
either ${\vec\Delta}_1 =0$, or  ${\vec\Delta}_2 =0$, i.e., SDW order is either $(0,\pi)$ or $(\pi,0)$. This is exactly the same SDW order
as observed  in the experiments.  If  $U_6+ U_8-U_7-U_4$ was negative,
$E^{ex}_{gr}$  would be minimized when  $|{\vec\Delta}_1| =|{\vec\Delta}_2|$, in which case the SDW OPs of the two  sublattices would align orthogonal to each other. The spin configuration  for such state is shown in
Fig. \ref{fig2}(c).
Such orthogonal spin configuration has been found in the analysis of spin ordering in the two-orbital model.\cite{jose}

\subsubsection{Deviations from perfect nesting}

Consider  next what happens when we also
include  into consideration  the fact that electronic pockets are actually ellipses rather than circles, {\it i.e.}, the effective masses $m_x$ and $m_y$ are not equal, and  $\varepsilon^{\beta_1}_{{\bf k}} \neq \varepsilon^{\beta_2}_{{\bf k}}$. To continue with the analytical analysis, we assume that the ellipticity is small,
introduce $m_x = (1+\delta)m$ and $m_y = (1-\delta)m$, where  $\delta <<1$,
and compute the correction to the ground state energy to  second order in $\delta$.
Performing the same set of transformations as before, we find that, for a non-zero $\delta$,  Eq.(\ref{eq:n7}) has to be supplemented by
\bea
&&H^{(1)}_{4} =  2 \delta \sum_{\bf p} \frac{p^2_x-p^2_y}{2m} \left\{ \cos 2\theta \left[ \cos^2\psi e^\dagger_{a{\bf p}}  e_{a{\bf p}}  \right. \right.
\nonumber \\
&& \left. \left. + \sin^2\psi p^\dagger_{a{\bf p}}  p_{a{\bf p}}  -  d^\dagger_{a{\bf p}}  d_{a{\bf p}} - \sin\psi \cos \psi \left(p^\dagger_{a{\bf p}}  e_{a{\bf p}} +h.c. \right)\right] \right. \nonumber\\
&& \left. + \sin 2\theta \left[\cos \psi \left(p^\dagger_{a{\bf p}}d_{a{\bf p}} + h.c. \right) - \sin \psi \left(p^\dagger_{e{\bf p}}d_{a{\bf p}} + h.c. \right) \right] \right\} \nonumber\\
\label{eq:ellipt}
\eea
where the angles $\theta$ and $\psi$ are defined in the same way as before, and
the overall factor of  2  accounts for  spin degeneracy. For a perfect nesting $e_a$[$p_a$] states are all empty [occupied], and for small
$ \delta << \Delta$  ellipticity does not change this.

From Eq.(\ref{eq:ellipt}) we obtain two contributions to $E_{gr}$ of
order $\delta^2$. One comes from virtual transitions to non-occupied states and is negative.
Another comes from the change of the dispersion of the ungapped
$\varepsilon^{\beta_1}_{\bf k}$ in the presence of ellipticity, and is positive.
The negative contribution comes  from
non-diagonal terms in (\ref{eq:ellipt}) taken to second order.
Applying a standard second order quantum-mechanical perturbation theory we obtain
\bea
\lefteqn{E^{a,ellipt}_{gr} = - \delta^2 \sin^2{2 \theta} \times } && \nonumber\\
&& \sum_{\bf p} \left(\frac{p^2_x - p^2_y}{2m}\right)^2 \frac{\Delta^2}{2E_{\bf p}} \left[\frac{1}{(E_{\bf p}+ |\varepsilon_{\bf p}|)^2}
- \frac{1}{(2 E_{\bf p})^2}\right] + ... \nonumber\\
\label{su_1}
\eea
where $\varepsilon_{\bf p}$ and $E_{\bf p}$ are defined after Eq. (\ref{eq:n7}), and dots stand for the terms which do not depend on $\theta$ and do not break the degeneracy. We remind that $\cos \theta = |\vec{ \Delta}_1|/\Delta$ and $\sin \theta = |\vec{\Delta}_2|/\Delta$, so that $\sin^2 (2 \theta) = 4
|\vec{\Delta}_1|^2 |\vec{\Delta}_2|^2/\Delta^4$. Replacing the sum in Eq.(\ref{su_1})  by the integral and re-scaling,
we obtain for the energy per unit area
\beq
E^{a,ellipt}_{gr} = - \delta^2 \sin^2{2 \theta} \frac{m \nu^2}{16\pi} F\left(\frac{\mu}{\Delta}\right)
\label{su_2}
\eeq
where
\bea
\lefteqn{F(x) = }&& \nonumber\\
&& \frac{2}{x^2} \int_{-x}^\infty \frac{(y+x)^2 \,dy}{\sqrt{y^2+1}} \left[\frac{1}{(\sqrt{y^2+1} +|y|)^2} - \frac{1}{4 (y^2+1)}\right]. \nonumber\\
\label{su_3}
\eea
At large $x$, {\it i.e.}, at small $\Delta/\mu$,
  expected  within the
 itinerant  description, $F(x\rightarrow \infty) \approx 1$, and $E^{a,ellipt}_{gr}$ becomes
\beq
E^{a,ellipt}_{gr} = - \delta^2 \sin^2{2 \theta} \frac{m \mu^2}{16\pi}.
\label{su_4}
\eeq
Another contribution to $E_{gr}$ comes from the diagonal term in Eq.(\ref{eq:ellipt}) and is related to $\delta-$induced change in the dispersion of  $d-$ fermions, which are not gapped by SDW. Adding $- \delta (p^2_x-p^2_y)/m$ term from
Eq.(\ref{eq:ellipt}) to the dispersion of a $d-$fermion and evaluating the energy of the occupied states, we obtain after simple algebra that a linear term in $\delta$ is canceled out, but $\delta^2$ term is finite and yields
\beq
E^{b,ellipt}_{gr} = + \delta^2 \sin^2{2 \theta} \frac{m \mu^2}{8\pi}
\label{su_5}
\eeq
Comparing the two contributions, we find that
$E^{b,ellipt}_{gr}$  is two times larger than $E^{a,ellipt}_{gr}$.  Adding the
two terms and expressing $\sin^2{2 \theta}$ in terms of $\Delta_1$ and $\Delta_2$, we obtain
\bea
E_{gr}^{ellipt} = C |\vec{\Delta}_1|^2 |\vec{\Delta}_2|^2,  ~~C =\delta^2 \frac{m \mu^2}{4 \pi \Delta^4}
\label{su_6}
\eea
We see that $C$ is positive, i.e., the correction due to ellipticity of electron pockets breaks the degeneracy and selects either $(0,\pi)$ or $(\pi,0)$ state.
This is again the same selection as one needs for consistency with the experiments. We found it quite  remarkable that ellipticity introduces effective interaction between two SDW OPs which, for $\Delta << \mu$,
leads to the same selection of the ground state SDW order as the direct interaction between the two electron pockets.

For larger $\Delta/\mu$, $F(x)$ increases, eventually to
$F(x << 1) = (1/2) (0.25/x)^2$, and $E^{a,ellipt}_{gr}$
 becomes larger than $E^{b,ellipt}_{gr}$.  The sign change of
$E^{ellipt}_{gr}$ occurs, however,
at quite large $\Delta \geq \mu/2$ which very likely
is not realized in the pnictides.

\subsubsection{Selection of the SDW order in the itinerant $J_1-J_2$ model.}
\label{subsubsec:4}

We next show that  the selection of the $(0, \pi)$ and $(\pi, 0)$ states
in the itinerant model occurs only if  fermion-fermion interactions
are conventional
charge-charge interactions rather than spin-spin interactions (the vertices
contain spin $\delta-$functions rather than $\sigma$-matrices).
For briefty, we consider the
itinerant $J_1-J_2$ model with spin-spin interaction
and show that this model also possesses $O(6)$ symmetry at the mean-field level, but the magnetic order selected by quartic terms is different from $(0,\pi)$.

The itinerant $J_1-J_2$  model is described by
\bea
&&H^{J_1-J_2} = \sum {\vec S}({\bf p}){\vec S}(-{\bf p}) \times  \nonumber \\
&& \left[J_1 \left(\cos p_x + \cos p_y) + 2 J_2 \cos p_x \cos p_y\right)\right]
\label{eq:n6}
\eea
where ${\vec S}({\bf p}) = (1/2) \sum_{p_1} a^\dagger_{{\bf p}_1 \alpha} {\vec \sigma}_{\alpha \beta} a_{{\bf p}_1+{\bf p} \beta}$, where $a$ are fermionic operators which can be either holes, when ${\bf p}_1$ is close to $(0,0)$,
or electrons, when ${\bf p}_1$ is close to $(0,\pi)$ or $(\pi,0)$.
The kinetic energy term is the same as in Eq. (\ref{eqH}).

We focus on  momenta ${\bf p}$ in (\ref{eq:n6})
 near $(0,0)$, $(0,\pi)$, $(\pi,0)$ and $(\pi,\pi)$,
re-write the Hamiltonian in terms of hole and electron operators,
 and  introduce the same SDW vector OPs as before:
$\Delta^z_1 = - J_{SDW}\sum_{\bf p} \langle \alpha^\dag_{1\mathbf{p}\uparrow} \beta_{1\mathbf{p} \uparrow}  \rangle $  and
$ \Delta_{2}^{z(x)}= - J_{SDW}
\sum_{\bf p} \langle \alpha^\dag_{1\mathbf{p}\uparrow} \beta_{2\mathbf{p} \uparrow(\downarrow)}  \rangle$, where $J_{SDW} =
 \left(4 J_2 -J_1\right)$.
  Keeping first only the
interactions between $\alpha$ and $\beta$ fermions and performing the same
mean-field decoupling of the 4-fermion terms as
in the previous section, we find that self-consistency equations for  ${\vec \Delta}_1$  and  ${\vec \Delta}_2$ are again identical and only specify the value of  $\Delta^2 = |{\vec \Delta}_1|^2 + |{\vec \Delta}_2|^2$:
\beq
1 =  \frac{J_{SDW}}{2N}
 \sum_{\bf p} \frac{1}{\sqrt{\left(\varepsilon^-_{\bf p}\right)^{2}+\Delta^{2}}}.
\label{eq:n2_1}
\eeq
The solution for $\Delta$ exists for  $J_2 > J_1/4$, {\it i.e.} for large
 $J_2$ SDW order is antiferromagnetic  along the diagonals.
The condition $J_2 > J_1/4$ is similar to $J_2 > J_1/2$ for a classical $J_1-J_2$ model of localized spins (the conditions in itinerant and localized models do not have to
be exactly the same, indeed).

The degeneracy between different SDW states with the same $\Delta^2$ is again
broken once we include interactions between electron pockets. Such interactions are generated by $J_1$ and $J_2$ terms taken at momenta $p = (0,0)$ and $(\pi,\pi)$.
Re-expressing $ \sum {\vec S}({\bf p}){\vec S}(-{\bf p})$ in terms of fermions
we obtain the quartic interaction between $\beta_1$ and $\beta_2$ fermions
 in the same form as in  Eq. (\ref{su_8}), with the coefficients
\bea
&& U^s_{6}  = J_1 - 3 J_2, ~ U^s_7 = -J_1 - 3 J_2, \nonumber \\
&& U^s_8 = 3 (J_1 - J_2), U^s_4 = -3 (J_1 + J_2).
\label{nn_1}
\eea
Substituting these coefficients into (\ref{energy}), we obtain
\beq
E^{J_1-J_2}_{gr} =  4 A^2 \left[4 J_1 \frac{|\vec{\Delta}_1|^2 |\vec{\Delta}_2|^2}{\Delta^4} -  \left(J_1 + 3 J_2 \right) \frac{\left(\vec{\Delta}_1 \cdot \vec{\Delta}_2 \right)^2}{\Delta^4}\right]
\label{energy_1}
\eeq
where $A>0$ is defined after Eq. (\ref{energy}).
Comparing this form with Eq. (\ref{energy}) we see that now the pre-factor for
$ \left(\vec{\Delta}_1 \cdot \vec{\Delta}_2 \right)^2$ term is negative, i.e., the energy  is lowered when
${\vec \Delta}_1$ and ${\vec \Delta}_2$ are parallel. The terms $|\vec{\Delta}_1|^2 |\vec{\Delta}_2|^2$ and $\left(\vec{\Delta}_1 \cdot \vec{\Delta}_2 \right)^2$ are then equal, and from (\ref{energy_1}) we obtain
 \beq
E^{J_1-J_2}_{gr} = -12  A^2 \left(J_2 - J_1\right)
\frac{|\vec{\Delta}_1|^2 |\vec{\Delta}_2|^2}{\Delta^4}
\label{energy_11}
\eeq
We see that the state with  ${\vec \Delta}_1 = 0$ or ${\vec \Delta}_2 =0$, which has $(0,\pi)$ or $(\pi,0)$ order, is only favored in the
 range $J_1/4 < J_2 < J_2$. For larger $J_2$, the energy is minimized when
${\vec \Delta}_1 = \pm {\vec \Delta}_2$.  The corresponding SDW state has
antiferromagnetic order for spins in one sublattice,
but no SDW order for spins in the other sublattice (see Fig. \ref{fig2}(d)).
Such state has been identified as one of possible candidates for the magnetic ground state in the generic analysis of the Landau theory for a two-component
order parameter.~\cite{jose}

We see therefore that, when $J_2$ interaction dominates, the spin-spin interaction between the two electron bands selects different SDW order from the case when the interaction  is in the charge channel.

\section{SDW order in four-band model}
\label{sec:3a}

So far we found that the stripe $(\pi,0)$ or $(0,\pi)$ order is selected
in the 3 band model (one hole and two electron FSs) with conventional
charge interactions. We now add the second hole pocket and check how
its inclusion affects the SDW order.

As we said in the Introduction, the second
hole FS is less coupled to electron FSs than the one that we already included
into the 3-band model. This is due to a combination of the two factors:  the difference in the  interactions $U_{SDW}$
and the difference in the degree of the overlap with the elliptic electron FSs. We consider both factors.

To begin, consider a model of two circular hole FSs and two circular
electron  FSs. Let's assume that  all FSs are of the same size, but
that there are two different SDW interactions
between hole and electron bands -- $U^{\{1\}}_{SDW}$ for one hole band,
 nd $U^{\{2\}}_{SDW}$ for the other. We introduce four SDW OPs:
${\vec \Delta}_{11}$, ${\vec \Delta}_{12}$,  ${\vec \Delta}_{21}$, and  ${\vec \Delta}_{22}$,  of which   ${\vec \Delta}_{11}$ and ${\vec \Delta}_{21}$ are with momentum
${\bf Q}_1$, and  ${\vec \Delta}_{12}$ and  ${\vec \Delta}_{22}$ are with momentum
${\bf Q}_2$. The OPs ${\vec \Delta}_{11}$ and  ${\vec \Delta}_{12}$ involve fermions from
the first hole band, while  ${\vec \Delta}_{21}$, and  ${\vec \Delta}_{22}$ involve
fermions from the second hole band.  Without loss of generality,
$\vec{\Delta}_{11}$ can be directed along $z$ axis, and  $\vec{\Delta}_{12}$ in the
$xz$ plane, but the directions of  $\vec{\Delta}_{21}$ and  $\vec{\Delta}_{22}$ can be arbitrary in 3D space. To simplify the  discussion, we assume that the SDW configuration is coplanar,
and set $\vec{\Delta}_{11}$ and  $\vec{\Delta}_{21}$ to be
along $z$-axis  and
$\vec{\Delta}_{12}$ and $\vec{\Delta}_{22}$ to be along $x$ axis.
In explicit form, we then have, by analogy with Eq. (\ref{f_1_1}),
\bea
&&{\vec \Delta}_{11} = \Delta_{1}^{z}= - U^{\{1\}}_{SDW}
\sum_{\bf p} \langle \alpha^\dag_{1\mathbf{p}\uparrow} \beta_{1\mathbf{p} \uparrow}  \rangle \nonumber \\
&&{\vec \Delta}_{12} = \Delta_{1}^{x}= -
U^{\{1\}}_{SDW} \sum_{\bf p} \langle \alpha^\dag_{1\mathbf{p}\uparrow} \beta_{2\mathbf{p}\downarrow}  \rangle \nonumber \\
&&{\vec \Delta}_{21} = \Delta_{2}^{z}= - U^{\{2\}}_{SDW}
\sum_{\bf p} \langle \alpha^\dag_{2\mathbf{p}\uparrow} \beta_{1\mathbf{p} \uparrow}  \rangle \nonumber \\
&&{\vec \Delta}_{22} = \Delta_{2}^{x}= -
U^{\{2\}}_{SDW} \sum_{\bf p} \langle \alpha^\dag_{2\mathbf{p}\uparrow} \beta_{2\mathbf{p}\downarrow}  \rangle.
\label{f_1}
\eea
From now on the subindices 1 and 2 indicate SDW OPs associated with one or the other hole bands.

As  in previous section, we
first consider only the interactions between hole and electron states which contribute to the SDW order $(U_1$ and $U_3$ terms). Cecoupling 4-fermion terms using Eq. (\ref{f_1}), we obtain the quadratic Hamiltonian in the form
 $H^{\{2\}}_{eff} = H^{kin} + H^{\{2\}}_{\alpha_1 \beta} +
^{\{2\}}_{\alpha_2 \beta}$, where
\bwt
\begin{eqnarray}
\label{eqH_1}
&&H^{kin} =  \sum_{\mathbf{p}, \sigma,i=1,2}  \varepsilon_p
 \left[ \alpha_{i \mathbf{p}  \sigma}^\dag \alpha_{i\mathbf{p} \sigma} -
 \beta_{i\mathbf{p}  \sigma}^\dag \beta_{i\mathbf{p} \sigma}\right] \\
&&H^{\{2\}}_{\alpha_1 \beta} =  -  \sum_{\mathbf{p}} \left[\alpha^\dag_{1 \mathbf{p} \uparrow} \left(\Delta^z_1 \beta_{1 \mathbf{p} \uparrow} +  \Delta^x_2 \beta_{2 \mathbf{p} \downarrow}\right)-
\alpha^\dag_{1 \mathbf{p} \downarrow} \left(\Delta^z_1 \beta_{1 \mathbf{p} \downarrow} -  \Delta^x_2 \beta_{2 \mathbf{p} \uparrow}\right)\right] + h.c. \nonumber \\
&&H^{\{2\}}_{\alpha_2 \beta} =  -   \sum_{\mathbf{p}} \left[\alpha^\dag_{2 \mathbf{p} \uparrow} \left(\Delta^z_3 \beta_{1 \mathbf{p} \uparrow} +  \Delta^x_4 \beta_{2 \mathbf{p} \downarrow}\right)-
\alpha^\dag_{2 \mathbf{p} \downarrow} \left(\Delta^z_3 \beta_{1 \mathbf{p} \downarrow} -  \Delta^x_4 \beta_{2 \mathbf{p} \uparrow}\right)\right] + h.c.
\end{eqnarray}
\ewt
The part involving $\alpha_1$ hole band and $\Delta^z_1$ and $\Delta^x_1$
 can be diagonalized in the same way as for the 3-band model, by introducing $\cos \theta = \Delta^z_1/\Delta_1,
\sin\theta = \Delta^x_1/\Delta_1$ where $\Delta_1 = \sqrt{(\Delta^z_1)^2 + (\Delta^x_1)^2}$ and rotating $\beta_{1,2}$ into
\bea
&&\beta_{1 \mathbf{p} \downarrow} = c_{b \mathbf{p}} \cos \theta
- d_{b \mathbf{p}} \sin \theta,\nonumber \\
&& ~~\beta_{2 \mathbf{p} \downarrow} =
c_{a \mathbf{p}} \sin \theta +  d_{a \mathbf{p}} \cos \theta, \nonumber \\
&&\beta_{1 \mathbf{p} \uparrow} = c_{a \mathbf{p}} \cos \theta
- d_{a \mathbf{p}} \sin \theta, \nonumber \\
&&\beta_{2 \mathbf{p} \uparrow} =
-c_{b \mathbf{p}} \sin \theta -  d_{b \mathbf{p}} \cos \theta,
\label{f_2}
\eea
Substituting this into (\ref{eqH_1}) we obtain
\beq
H^{\{2\}}_{\alpha_1 \beta} = - \Delta_1
 \sum_{\mathbf{p}} \left[\alpha^\dag_{1 \mathbf{p} \uparrow} c_{a \mathbf{p}} -
\alpha^\dag_{1 \mathbf{p} \downarrow} c_{b \mathbf{p}}\right]
\label{f_3}
\eeq
and $H^{kin} = H^{kin}_{\{\alpha_1, c\}} + H^{kin}_{\{\alpha_2, d\}}$, where
\bea
&& H^{kin}_{\{\alpha_1, c\}} = \nonumber \\
&& \sum_{\mathbf{p}}  \varepsilon_{\bf p}
 \left[ \alpha_{1 \mathbf{p} \uparrow}^\dag \alpha_{1\mathbf{p} \uparrow} +
 \alpha_{1 \mathbf{p} \downarrow}^\dag \alpha_{1\mathbf{p} \downarrow} -
 c_{a \mathbf{p}}^\dag c_{a \mathbf{p}}  - c_{b \mathbf{p}}^\dag c_{b \mathbf{p}}\right] \label{f_4a} \nonumber
 \\
 \\
&&H^{kin}_{\{\alpha_2, d\}} = \nonumber \\
&& \sum_{\mathbf{p}}  \varepsilon_{\bf p}
 \left[ \alpha_{2 \mathbf{p} \uparrow}^\dag \alpha_{2\mathbf{p} \uparrow} +
 \alpha_{2 \mathbf{p} \downarrow}^\dag \alpha_{2 \mathbf{p} \downarrow} -
 d_{a \mathbf{p}}^\dag d_{a \mathbf{p}}  - d_{b \mathbf{p}}^\dag d_{b \mathbf{p}}\right] \nonumber\\
\label{f_4}
\eea
The part $ H^{kin}_{\{\alpha_1, c\}} + H^{\{2\}}_{\alpha_1 \beta}$
 involves hole $\alpha_1$ operators and electron $c_{a,b}$ operators:
\bea
&&  H^{kin}_{\{\alpha_1, c\}} + H^{\{2\}}_{\alpha_1 \beta} = \nonumber \\
 && \sum_{\mathbf{p}}  \varepsilon_{\bf p}
 \left[ \alpha_{1 \mathbf{p} \uparrow}^\dag \alpha_{1\mathbf{p} \uparrow} +
 \alpha_{1 \mathbf{p} \downarrow}^\dag \alpha_{1\mathbf{p} \downarrow} -
 c_{a \mathbf{p}}^\dag c_{a \mathbf{p}}  - c_{b \mathbf{p}}^\dag c_{b \mathbf{p}}\right] \nonumber \\
&& - \Delta_1
 \sum_{\mathbf{p}} \left[\alpha^\dag_{1 \mathbf{p} \uparrow} c_{a \mathbf{p}} -
\alpha^\dag_{1 \mathbf{p} \downarrow} c_{b \mathbf{p}}\right]
\label{f_5_1}
\eea
Eq. (\ref{f_5_1}) can  be diagonalized by the same transformation as in Eqs. (\ref{f_5}) and (\ref{f_6}):
\bea
&&\alpha_{1{\bf p}\uparrow} =
 p_{a {\bf p}} \cos \psi + e_{a {\bf p}} \sin \psi, \nonumber \\
&&~c_{a {\bf p}} =
e_{a {\bf p}} \cos \psi - p_{a {\bf p}} \sin \psi, \nonumber \\
&&\alpha_{1{\bf p}\downarrow}  =  p_{b {\bf p}} \cos \tilde{\psi} +
e_{b {\bf p}} \sin \tilde{\psi}, \nonumber \\
&&c_{b{\bf p}} = e_{b {\bf p}} \cos \tilde{\psi} - p_{b {\bf p}}
\sin \tilde{\psi},
\label{f_7}
\eea
\\
where, as before, $\tilde{\psi}=\frac{\pi}{2} +\psi$  and
$\cos^2 \psi, ~\sin^2 \psi =
\frac{1}{2} \left[ 1 \pm \frac{\varepsilon_{\bf
p}}{\sqrt{\left(\varepsilon_{\bf p}\right)^{2}+\Delta^{2}_1}} \right]$.
Substituting, we obtain
\bea
&& H^{kin}_{\{\alpha_1, c\}} + H^{\{2\}}_{\alpha_1 \beta} = \nonumber \\
&&
\sum_{p} E_{1, {\bf p}} \left[ e^\dagger_{a{\bf p}}  e_{a{\bf p}} + p^{\dagger}_{b{\bf p}}  p_{b{\bf p}} - e^\dagger_{b{\bf p}}  e_{b{\bf p}} - p^{\dagger}_{a{\bf p}}  p_{a{\bf p}}\right],
\label{f_9}
\eea
where
$E_{1, \bf p} =  \pm \sqrt{\left(\varepsilon_{\bf p}\right)^2+\Delta^2_1}$.
This result is similar to Eq. (\ref{eq:n7}).
The self-consistent equation for the gap $\Delta_1$ is
\beq
1 = \frac{
U^{\{1\}}_{SDW}}{2N} \sum_{\bf p} \frac{1}{\sqrt{\left(\varepsilon^-_{\bf p}\right)^{2}+\Delta^{2}_1}}.
\label{f_10}
\eeq
 The remaining part of the Hamiltonian involves holes from $\alpha_2$ band
 and reduces to
\bwt
 \bea
&& H^{kin}_{\{\alpha_2, d\}} + H^{\{2\}}_{\alpha_2 \beta} =
\sum_{\mathbf{p}}  \varepsilon_{\bf p}
 \left[ \alpha_{2 \mathbf{p}  \uparrow}^\dag \alpha_{2\mathbf{p} \uparrow}
+ \alpha_{2 \mathbf{p}  \downarrow}^\dag \alpha_{2\mathbf{p} \downarrow}
 - d_{a\mathbf{p}}^\dag d_{a\mathbf{p}} -  d_{b\mathbf{p}}^\dag d_{b\mathbf{p}} \right] \nonumber \\
&& -\Delta_2 \left[\alpha_{2 \mathbf{p}  \uparrow}^\dag \left(c_{a,\mathbf{p}} \cos ({\tilde \theta} -\theta) + d_{a,\mathbf{p}}  \sin ({\tilde \theta} -\theta)\right) - \alpha_{2 \mathbf{p}  \downarrow}^\dag
\left(c_{a,\mathbf{p}} \cos ({\tilde \theta} -\theta) + d_{a,\mathbf{p}}  \sin ({\tilde \theta} -\theta)\right) \right]
\label{f_11}
\eea
\ewt
where we introduced
$\cos {\tilde \theta} = \Delta^z_2/\Delta_2,
\sin{\tilde \theta} = \Delta^x_2/\Delta_2$ and $\Delta_2 = \sqrt{(\Delta^z_2)^2 + (\Delta^x_2)^2}$.

This part of the Hamiltonian decouples from the one which we already
diagonalized if we eliminate the terms with $c-$operators. This
will be the case if we choose ${\tilde \theta} = \pi/2 + \theta$. Then Eq. (\ref{f_11}) becomes the quadratic form of hole $\alpha_2$ operators and electron $d_{a,b}$ operators neither of which is present in the quadratic form of Eq. (\ref{f_5_1}).
Diagonalizing the quadratic form in Eq. (\ref{f_11}),
we obtain, in terms of new operators
${\tilde e}_{a,b}$ and ${\tilde p}_{a,b}$,
\bea
&& H^{kin}_{\{\alpha_2, d\}} + H^{\{2\}}_{\alpha_2 \beta} = \nonumber \\
&&\sum_{p} E_{{2, \bf p}} \left[ {\tilde e}^\dagger_{a{\bf p}}  {\tilde e}_{a{\bf p}} + {\tilde p}^{\dagger}_{b{\bf p}}  {\tilde p}_{b{\bf p}} - {\tilde e}^\dagger_{b{\bf p}}  {\tilde e}_{b{\bf p}} - {\tilde p}^{\dagger}_{a{\bf p}}  {\tilde p}_{a{\bf p}}\right],
\label{f_12}
\eea
where
$E_{2, \bf p} =  \pm \sqrt{\left(\varepsilon_{\bf p}\right)^2+\Delta^2_2}$,
The self-consistent equation for the gap $\Delta_2$ is
\beq
1 = \frac{
U^{\{2\}}_{SDW}}{2N} \sum_{\bf p} \frac{1}{\sqrt{\left(\varepsilon^-_{\bf p}\right)^{2}+\Delta^{2}_2}}.
\label{f_14}
\eeq
We see that, with the choice of ${\tilde \theta} = \pi/2 - \theta$, we decoupled the quadratic form into two parts, one for $\Delta_1$, another for $\Delta_2$.
 One can easily check that for both parts it is energetically favorable
 to have a state with non-zero $\Delta$ once self-consistent equations for
 $\Delta_{1,2}$ have solutions. This implies that the energy is further
reduced  if, in addition to creating $\Delta^z_1 = \Delta_1 \cos \theta$ and $\Delta^x_1 = \Delta_1 \sin \theta$, one also creates $\Delta^z_2 = - \Delta_2 \sin \theta$ and $\Delta^x_2 = \Delta_2 \cos \theta$.
Furthermore, for equal-size circular FSs, the solutions
for $\Delta_1$ and $\Delta_2$ exist for any $U^{\{1\}}_{SDW}$ and $U^{\{2\}}_{SDW}$.
The resulting fermionic excitations are fully gapped, i.e., within 4-band model of equal size circular hole and electron pockets, an SDW state is an insulator.

We now verify whether such insulating state is consistent with the observed $(0,\pi)$ or $(\pi,0)$ order.
Because the angle $\theta$ disappears from the quadratic forms in Eq. (\ref{f_5_1})
and Eq. (\ref{f_11}) (for ${\tilde \theta} = \pi/2 - \theta$), the ground state is again degenerate. The degenerate SDW OP manifold is given by
\bea
{\vec S} ({\bf R}) &\propto&
 \vec{n}_z  \left(\Delta_1 \cos \theta - \Delta_2 \sin \theta \right)
 e^{i{\bf Q}_1 {\bf R}} \nonumber \\
&&+ \vec{n}_x \left(\Delta_1 \sin \theta + \Delta_2 \cos \theta\right)  e^{i{\bf Q}_2 {\bf R}}
\label{f_15}
\eea
This degenerate set does contain
$(0,\pi)$ and $(\pi,0)$ states with only one ordering vector (either ${\bf Q}_1$ or ${\bf Q}_2$). These states are obtained if we set $\tan \theta = \Delta_1/\Delta_2$ or $\tan \theta = - \Delta_2/\Delta_1$. The issue then is whether such
states  are selected by other interactions in the same spirit as $(0,\pi)$ or $(\pi,0)$ orders were selected in the 3-band model. We argue that they are not.  The argument is that $(0,\pi)$ and $(\pi,0)$ ordered states in the 4-band model are obtained by choosing $\theta$ such that
${\vec \Delta}_1$  and ${\vec \Delta}_2$  have components with both momenta,
${\bf Q}_1$ and ${\bf Q}_2$. The total spin component with either   ${\bf Q}_1$ or ${\bf Q}_2$ vanishes because of cancelation between $\Delta_1$ and $\Delta_2$ components.
Indeed, for $U^{\{2\}}_{SDW} << U^{\{1\}}_{SDW}$
$\theta$ is close to zero or to $\pi/2$, but it is not equal to either of these values. Meanwhile, when we considered  lifting of the degeneracy in the 3-band model, we found that the interactions between $\beta_1$ and $\beta_2$ electrons and  the ellipticity give rise to the $\theta$ dependence of $E_{gr}$ in the form $E_{gr} (\theta)
 = E_0 + E_1 \sin^2 (2\theta)$ such that $E_{gr} (\theta)$ has minima at $\theta$ exactly equal to  $0$ or to
$\pi/2$.
To verify what happens in the 4-band model, we extended those calculations to the case when $U^{\{2\}}_{SDW}$ is non-zero and analyzed the form of $E_{gr}
(\theta)$ perturbatively in $U^{\{2\}}_{SDW}/U^{\{1\}}_{SDW}$. We skip the details of the calculations  as they are similar to the ones for the 3 band model. We found that the minima in $E_{gr} (\theta)$ {\it do not} shift from
$\theta =0$ and $\theta = \pi/2$,
 at least at small $U^{\{2\}}_{SDW}/U^{\{1\}}_{SDW}$. The consequence is that, for the $4-band$ model, interactions which break the degeneracy of SDW manifold favor the states  which {\it are not} $(0,\pi)$ or $(\pi,0)$ states.
For example, when $\theta =0$, ${\vec S} ({\bf R}) \propto
\Delta_1  \vec{n}_z  e^{i{\bf Q}_1 {\bf R}}+
\Delta_2  \vec{n}_x  e^{i{\bf Q}_2 {\bf R}}$. Such SDW state is a two-sublattice structure with equal magnitudes of the order parameters on the two sublattices, but with a non-collinear orientation of

The outcome of this study is that $(0,\pi)$ or $(\pi,0)$ order can only
be preserved if $\Delta_2$ is strictly zero, i.e., the second hole band is not involved in the SDW mixing.
Only then interactions and ellipticity select $(0,\pi)$ or $(\pi,0)$ states.
Otherwise, the order necessary has both ${\bf Q}_1$ and ${\bf Q}_2$ components and the SDW OP has  modulations along both $x$ and $y$ directions.

This observation has an implication for the electronic structure. When
$\Delta_2 =0$, one hole band and one electron band are not involved in the SDW mixing, i.e., the system {\it remains a metal even when SDW mixing of the other two bands is strong}.  In other words, $(0,\pi)$ and $(\pi,0)$ SDW order necessary involves only one of two hole bands and only one
of the two  electron bands. The other hole and electron bands remain intact.

\begin{figure}
\includegraphics[angle=0,width=1.0\columnwidth]{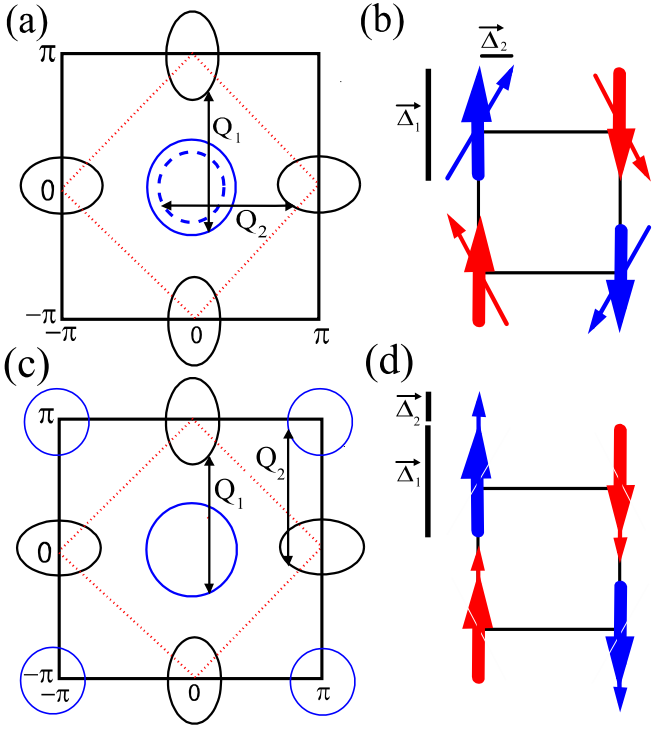}
\caption{(color online)
The changes to the magnetic structure introduced by the weak SDW coupling
 $\vec{\Delta}_2$ between the second hole band and the  electron bands.
Panels (a) and (b) are for the FS topology of the pnictides, panels
(c) and (d) are for the  FS topology of $t'-t$ model for $t\ll t'$.} \label{fig6}
\end{figure}

\begin{figure}
\includegraphics[angle=0,width=0.8\columnwidth]{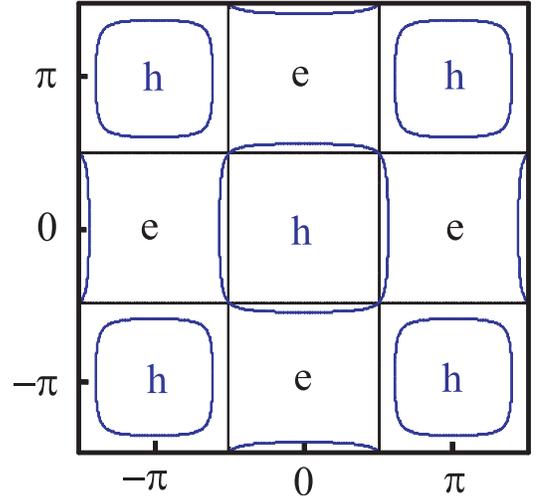}
\caption{(color online) The FS
for the tight-binding model $\varepsilon_{\bf p} = 4 t' \cos p_x \cos p_y + 2t\left(\cos p_x +\cos p_y\right)$ for $t=0$ (solid curves) and $t=0.5t'$ (blue curves).  $h$ and $e$ refer to the hole and electron states occupying the corresponding parts of the BZ. Observe that the hole bands are now centered at $(0,0)$ and $(\pi,\pi)$.
} \label{fig7}
\end{figure}

As we said,
for equal-size circular hole and electron pockets, there is a perfect nesting between both primary and secondary pairs of hole and electron states, and
$\Delta_2$ is non-zero for any non-zero $U^{\{2\}}_{SDW}$. This is, however, not the case when the two hole FSs are circles of unequal size, and the two electron FSs are ellipses. In this situation, re-doing the same calculations as above we obtain in general 4 different SDW bands described by
\bea
H_{SDW} & = & \sum_{a,b} \sum_{p} E_{1,2{\bf p}} \left( e^\dagger_{a,b{\bf p}}  e_{a,b{\bf p}} + p^{\dagger}_{b,a{\bf p}}  p_{b,a{\bf p}}\right) +
\nonumber\\
&&  E_{3,4{\bf p}} \left(e^\dagger_{a,b{\bf p}}  e_{a,b{\bf p}} + p^{\dagger}_{b,a{\bf p}}  p_{b,a{\bf p}}\right),
\label{eq:next}
\eea
Here
\beq
E_{1,2{\bf p}} = \frac{1}{2} \left(\varepsilon^{\alpha_1}_{\bf k} + \tilde{\varepsilon}_{\bf k}\right) \pm
 \frac12\sqrt{\left(\varepsilon^{\alpha_1}_{\bf k} - \tilde{\varepsilon}_{\bf k}\right)^2+4|\Delta_1|^2}
\label{f_16}
\eeq
and
\beq
E_{3,4{\bf p}} = \frac{1}{2} \left(\varepsilon^{\alpha_2}_{\bf k} + \tilde{\tilde{\varepsilon}}_{\bf k}\right) \pm \frac12 \sqrt{\left(\varepsilon^{\alpha_2}_{\bf k} - \tilde{\tilde{\varepsilon}}_{\bf k}\right)^2+4|\Delta_2|^2},
\label{f_17}
\eeq
where $\tilde{\varepsilon}_{\bf k}=\varepsilon^{\beta_1}_{\bf k} \cos^2 \theta +\varepsilon^{\beta_2}_{\bf k} \sin^2 \theta$
and $\tilde{\tilde{\varepsilon}}_{\bf k}=\varepsilon^{\beta_2}_{\bf k} \cos^2 \theta +\varepsilon^{\beta_1}_{\bf k} \sin^2 \theta$.
The self-consistent equations for the two gaps are
\bea
1 & = & U^{\{1\}}_{SDW} \sum_p \frac{n\left(E_{1{\bf p}}\right)-n\left(E_{2{\bf p}}\right)}{\sqrt{\left(\varepsilon^{\alpha_1}_{\bf k} - \tilde{\varepsilon}_{\bf k}\right)^2+4|\Delta_1|^2}}, \\
1 & = & U^{\{2\}}_{SDW} \sum_p \frac{n\left(E_{3{\bf p}}\right)-n\left(E_{4{\bf p}}\right)}{\sqrt{\left(\varepsilon^{\alpha_2}_{\bf k} - \tilde{\tilde{\varepsilon}}_{\bf k}\right)^2+4|\Delta_2|^2}}.
\label{eq:nn2}
\eea
Analyzing these equations, we found that, for a non-perfect nesting,
SDW magnetism is a threshold phenomenon, i.e., to obtain a non-zero $\Delta_1$ and $\Delta_2$, the interactions $U_{SDW}$ must exceed the thresholds.
We found that a hole band with a heavier mass, e.g.,
with a {\it larger FS},
is more strongly coupled to electron bands. This band then plays the role of the $\alpha_1$ band in our theory. Once the corresponding $U^{\{1\}}_{SDW}$ exceeds the threshold value $U^{\{1\}}_{cr}$, the system develops an SDW order ($\Delta_1 \neq 0$) This order, as we know, is a stripe $(0,\pi)$ or $(\pi,0)$ order.
For the smaller-size hole
band ($\alpha_2$ band in our terminology), SDW order with $\Delta_2 \neq 0$
emerges only when $U^{\{2\}}_{SDW}$ exceeds a larger threshold $U^{\{2\}}_{cr} > U^{\{1\}}_{cr}$.  Once this happens, SDW order acquires both ${\bf Q}_1$ and ${\bf Q}_2$ components. This is illustrated in Fig.\ref{fig6}(a)-(b).
We see therefore that the regime consistent with the experiments is $U^{\{1\}}_{SDW} > U^{\{1\}}_{cr}$ and $U^{\{2\}}_{SDW} < U^{\{2\}}_{cr}$.
This regime is quite realistic given that there is quite sizable difference between the areas of the two hole pockets.

\subsection{$t-t'-U$ model}
\label{sec:t-t'}

In simple terms, our result that the
stripe order is only consistent with $\Delta_1 \neq 0$, $\Delta_2 0$
is ultimately related to the geometry of the FS, namely to the fact
that both hole FSs are located around the $\Gamma$ point. Indeed, suppose that
one hole and one electron FSs are mixed into an SDW state with the momenta
between these FSs (say $({\bf Q}_1 = (0,\pi)$). For strong interaction, the
SDW excitations are gapped and effectively decouple from the other
hole and electron bands.  These two bands can also mix into an
SDW state, however, because the second hole band is centered at $(0,0)$,
it is separated from the second electron band by ${\bf Q}_2 = (\pi,0)$, i.e.,
the second SDW order necessary has momentum $(\pi,0)$, and
the stripe order gets broken.

The SDW order would be different if the second FS was centered at
$(\pi,\pi)$ because then the remaining hole and electron bands would be
separated by the same ${\bf Q}$ as the first pair of  bands, and the stripe
configuration would  not be broken by the emergence of the second SDW order.
 This is illustrated in Fig.\ref{fig6}(c)-(d).

Such FS geometry is realized in the half-filled $t-t'-U$ model described by
\beq
H = t \sum_{i, \delta_1 \sigma} c^\dag_{i \sigma}   c_{i + \delta_1 \sigma} + t'  \sum_{i, \delta_2 \sigma} c^\dag_{i \sigma}   c_{i + \delta_2 \sigma} + U \sum_i n_{i \uparrow} n_{i \downarrow}
\label{f_18}
\eeq
where $\delta_1$ and $\delta_1$ are distances to nearest and next-nearest neighbors, and $n_{i a} = c^\dag_{i a} c_{i a}$.  In momentum space, the dispersion
$\varepsilon_{\bf p} = 2t (\cos{p_x} + \cos{p_y}) + 4 t' \cos{p_x} \cos{p_y} - \mu$
has maxima at $(0,0)$ and $(\pi,\pi)$ and isotropic quadratic hole-like dispersion around these points (with non-equal masses near $(0,0)$ and $(\pi,\pi)$, when $t \neq 0$), and minima at $(0,\pi)$ and $(\pi,0)$ and elliptical electron-like dispersion near these points. We illustrate this in Fig. \ref{fig7}. Such
band structure  is topologically equivalent to the one shown in
Fig.\ref{fig6}(c).

At large $U$, the $t-t'-U$
model reduces by standard manipulations to the $J_1-J_2$ Heisenberg spin model. The electronic states in this model are all gapped, and the
SDW OP is degenerate at the mean-field level.
We analyzed SDW order in the $t-t'-U$ model for arbitrary $U$ within the same computational scheme as before and found that $(0,\pi)$ and $(\pi,0)$ states minimize the energy at the mean-field level because both $\vec{\Delta}_1$ and $\vec{\Delta}_2$ appear with the same momentum.

It is instructive to consider the discrepancy between our 4-band model and
$t-t'-U$ model is some more detail. We remind that in our model,
  ellipticity of electron pockets and interactions between them
select $(0,\pi)$ and $(\pi,0)$ states already at the mean-field level.
 We show below  that in  $t-t'-U$ model  there is no selection of a particular SDW order at the mean-field level even when $t \neq 0$ and electron dispersion is elliptical. In this situation, $(0,\pi)$ and $(\pi,0)$ states
 remain degenerate  with infinite number of other two-sublattice states.
  Beyond mean-field level,  quantum fluctuations select
 $(0,\pi)$ or $(\pi,0)$ order in the $J_1-J_2$ model.
  We haven't check the selection of the SDW order at smaller $U$,
 but it is likely that $(0,\pi)$ and $(\pi,0)$ states
 are  selected for all values of $U$.

To demonstrate that mean-field SDW OP in $t-t-U$ model remains degenerate,
 consider this model at small $t$ and introduce  two SDW order parameters ${\vec \Delta}_1 =- (U/2N) \sum_{\bf k} \langle c^\dag_{{\bf k},\alpha} {\bf \sigma}_{\alpha \beta}  c _{{\bf k}+ {\bf Q}_1,\beta} \rangle$ with momentum ${\bf Q}_1$,  and
${\vec \Delta}_2 =-(U/2N) \sum_{\bf k} \langle c^\dag_{{\bf k},\alpha} {\bf \sigma}_{\alpha \beta}  c _{{\bf k}+ {\bf Q}_2,\beta}\rangle$ with momentum ${\bf Q}_2$.
 We introduce four fermionic operators with momenta near $(0,0)$, $(0,\pi)$, $(\pi,0)$, and $(\pi,\pi)$, and re-express $t-t'-U$ model as the
 model of holes and electrons.
 We skip computational details, which are not different from what we presented in previous Sections, and only cite the results.
 The extra terms in the quadratic form come from the interactions
 between holes and electrons. For $t=0$, all 4 dispersions are equal
(up to the overall sign), and  the Hamiltonian is diagonalized in the same way as in Sec. \ref{sec:2a}. Not surprising, the ground state energy depends
only on $(\vec{\Delta_1})^2 + (\vec{\Delta_2})^2$, i.e., there exists a degenerate manifold of SDW OPs.

Consider next the effect of the interactions between the electron bands.
There are four effective electron-electron interactions -- the analogs of $U_6, ~U_7,~U_8$ and $U_4$ in (\ref{eq:n4}).  They all originate from the same $U$ term and have the same prefactors.  According to Eq. (\ref{energy}), the extra term in the ground state energy from electron-electron interaction only has
$(\vec{\Delta}_1 \cdot \vec{\Delta}_2)^2$ part:
\beq
E^{ex}_{gr} =  4 A^2 U \frac{\left(\vec{\Delta}_1 \cdot \vec{\Delta}_2 \right)^2} {\Delta^4}
\label{energy_1_1}
\eeq
This term orders ${\vec \Delta}_1$ and ${\vec \Delta}_2$ perpendicular to each other, what makes the magnitudes of the two sublattice OPs ${\vec \Delta}_1 + {\vec \Delta}_2$ and  ${\vec \Delta}_1 - {\vec \Delta}_2$ equal. But in the absence of $|\vec{\Delta_1}|^2 \cdot |\vec{\Delta_2}|^2$ term, the angle between the two sublattices remains arbitrary.

Consider next the effect of  ellipticity of electron pockets. In our model, we remind, ellipticity gave rise to two contributions to $E_{gr}$, both scale as
$|\vec{\Delta_1}|^2 \cdot |\vec{\Delta_2}|^2$. The two contributions were of opposite sign, but were not equal.  In $t-t'-U$ model, the situation is very similar -- there are again two contributions to $E_{gr}$: one is the second-order contribution from non-diagonal terms in the Hamiltonian, induced by $t$,  another
 comes from the change of the dispersion of the diagonal terms. The contribution from non-diagonal terms is given by
\bea
\lefteqn{E^{a, ellipt}_{gr} = 8 t^2 \frac{|\vec{\Delta_1}|^2 |\vec{\Delta_2}|^2}{\Delta^4} \times} && \nonumber \\
&& \sum_{\bf p} \frac{\left(\cos^2{p_x} + \cos^2{p_y}\right) \Delta^2}{\left[\Delta^2 + 16 (t')^2 \cos^2{p_x} \cos^2{p_y}\right]^{3/2}} + ...
\label{f_19}
\eea
where, as before, the dots stand for the terms which depend only on $\Delta$.
The contribution from the change of the dispersion of the diagonal terms in the Hamiltonian is
\bwt
\beq
E^{b, ellipt}_{gr} = 2 \sum_p \left[\varepsilon_{c1} + \varepsilon_{c2} + \varepsilon_{fa} + \varepsilon_{fb} - \sqrt{\left(\frac{\varepsilon_{c1} + \varepsilon_{fa}}{2}\right)^2 + \Delta^2} - \sqrt{\left(\frac{\varepsilon_{c2} + \varepsilon_{fb}}{2}\right)^2 + \Delta^2}\right] \quad,
\label{f_20}
\eeq
\ewt
where
\bea
&&\varepsilon_{c1} = 4t' \cos{p_x} \cos{p_y} + 2t \left( \cos{p_x} + \cos{p_y}\right), \nonumber \\
&&\varepsilon_{c1} = 4t' \cos{p_x} \cos{p_y} - 2t \left( \cos{p_x} + \cos{p_y}\right), \nonumber \\
&&\varepsilon_{fa} =  - 4t' \cos{p_x} \cos{p_y} + 2t \left( \cos{p_x} + \cos{p_y}\right) \left(\frac{{\vec \Delta}^2_1 - {\vec \Delta}^2_2}{\Delta^2}\right) \nonumber \\
&&\varepsilon_{fb} =  - 4t' \cos{p_x} \cos{p_y} - 2t \left( \cos{p_x} + \cos{p_y}\right) \left(\frac{{\vec \Delta}^2_1 - {\vec \Delta}^2_2}{\Delta^2}\right)\nonumber\\
\label{f_21}
\eea
Substituting these energies into (\ref{f_20}) and expanding in $t$, we find
\bea
\lefteqn{E^{b, ellipt}_{gr} = - 8 t^2 \frac{|\vec{\Delta_1}|^2 |\vec{\Delta_2}|^2}{\Delta^4}}&& \nonumber \\
&&  \sum_{\bf p} \frac{\left(\cos^2{p_x} + \cos^2{p_y}\right)) \Delta^2}{\left[\Delta^2 + 16 (t')^2
 \cos^2{p_x} \cos^2{p_y}\right]^{3/2}} + ...\nonumber \\
\label{f_22}
\eea
Comparing (\ref{f_19}) and (\ref{f_22}), we find that
$E^{a, ellipt}_{gr} + E^{b, ellipt}_{gr} =0$, i.e., ellipticity of electron bands in $t-t'-U$ model does not give rise to a selection of a particular SDW OP
 from the degenerate manifold.  This result very likely holds for arbitrary
$t/t'$, as long as the ground state manifold consists of states with antiferromagnetic order along the diagonals.

\section{Electronic structure of the SDW state}
\label{sec:5}

\begin{figure}[t]
\includegraphics[angle=0,width=1\columnwidth]{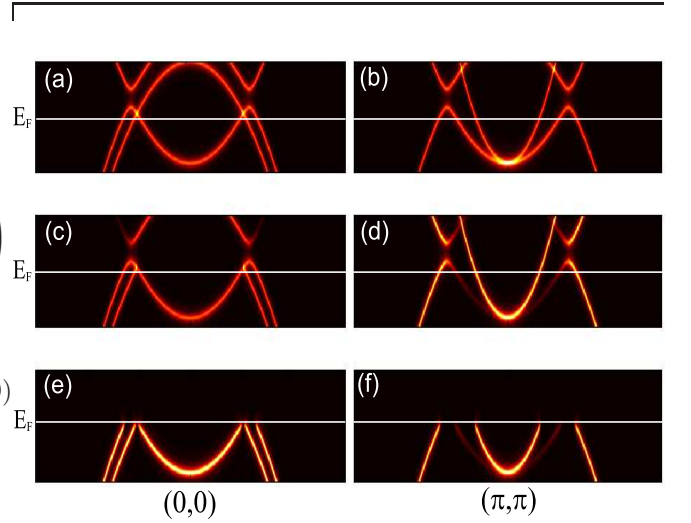}
\caption{(color online) Calculated image of the electronic structure in the folded BZ  in the SDW phase. Left panel -- near $\Gamma = (0,0)$, right panel -- near $M = (\pi,\pi)$.
In figures  (a) and (b) we show the  dispersion, in figures (c) and (d)
the spectral function $Im G({\bf k},\omega) = \sum_i Im G_i$, and in figures
(e) and (f) ARPES intensity, i.e., the product $Im G({\bf k},\omega) \times f(E_{\bf k})$. The direction of momenta in all figures is along $k_x =k_y$ in the folded BZ. We  directed SDW order parameter
along $z$ axis and set $\Delta^{z}_1 \neq 0$, $\Delta_2 = \Delta_3 = \Delta_4 =0$ ($(0,\pi)$ order).  For definiteness we set $\Delta^z_1 = \mu/4$, $\mu_1 =(3/2) \mu$, $\mu_2 = (3/2)\mu$, and  $m_1=m_2=m_y=2 m_x = 1$.
For numerical purposes we added the small damping constant $\delta = 0.04 \mu$.}
\label{fig4}
\end{figure}

As we said earlier, when all four Fermi surfaces are circles of equal size,
all 4 Fermi surfaces are mixed by the SDW and  are gapped.
However, as soon as the two hole pockets become non-equal,
there is a range of the interactions when one hole pocket and one of
the two electron pockets are mixed by  SDW, but the other hole and electron
pockets remain intact. As a consequence, in the folded BZ, the electronic structure contains one hole and one electron band mixed by SDW order and visible both near $(0,0)$ and $(\pi,\pi)$,  one hole band visible only near  $(0,0)$, and one electron band visible only near  $(\pi,\pi)$.   The only
effect of SDW ordering on the non-mixed excitations is the  $\Delta$-dependence of the intensity via coherence factors. Besides, for moderately strong interactions, as in the pnictides, $\Delta_1$ is not too large, and the
SDW-mixed dispersions  form the new FSs~\cite{parker}. The net result is
 the presence of three bands  near the $\Gamma$ point, one of which (the hole band) is not modified by SDW,
and another three bands  close to the $(\pi,\pi)$ point, one of which (the electron band) is not modified by SDW (see Figs.\ref{fig4}(a)-(b)).

The spectral functions $\sum_i Im G_i$ are peaked at quasiparticle energies
 but are also sensitive, via the SDW coherence factors,
to the interplay between the effective masses of the bare hole and electron dispersions.
For example, for two equivalent hole pockets ($m_1 =m_2$) and $m_x < m_1 = m_y$,
 the  spectral weight of
 ARPES intensity  $\sum_i n_F(E_{{\bf k}i})\times \mbox{Im} G_i$ ($n_F(E)$ is a Fermi function), which we plot in
Figs.\ref{fig4}(e)-(f), is the largest for
two hole-like bands centered around $\Gamma$-point, and for one electron band and hole ``half-bands'' (hole blades)
around $(\pi,\pi)$-point. This theoretical ARPES intensity is
quite consistent with the experimental
data
from Refs. \cite{kondo,zhou,dresden_arpes}: the experiments also  show
two hole dispersions near $\Gamma$ point,  and  one electron band
  and two hole
``half-bands''  near $(\pi,\pi)$. There is
some evidence~\cite{zhou} of the third hole dispersion near $(0,0)$, but this one likely emerges from the fifth band which we did not consider here.

\begin{figure}[t]
\includegraphics[angle=0,width=1\columnwidth]{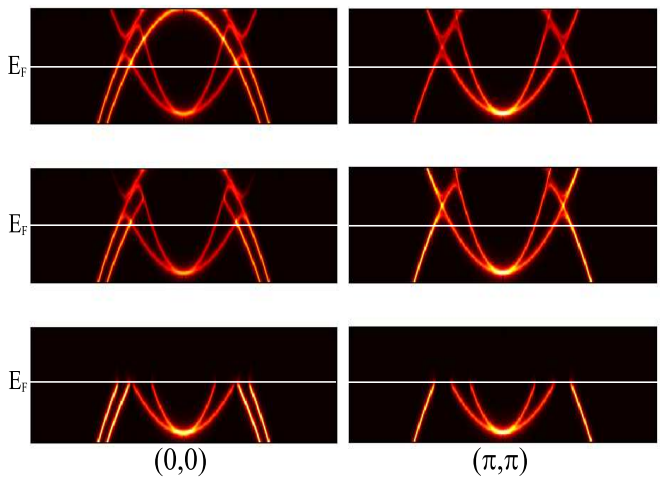}
\caption{(color online)
Same as in Fig.\protect\ref{fig4} but for a multi-domain sample with
equal distribution of the domains with $(\pi,0)$ and ($0,\pi$) SDW orders.}
\label{fig5}
\end{figure}

Our results are also consistent with recent observations
of ``anisotropic Dirac cones''.\cite{richard,suchitra}
 Indeed,  the bands mixed in the SDW state form small hole
pockets near both $\Gamma$-and $M$-points of the BZ, and the dispersion around these pockets is almost linear close to the
Fermi level and also anisotropic between  $k_x$ and $k_y$ due to initial anisotropy of the elliptic band involved in the SDW mixing. However, despite of
a visual similarity with the Dirac cone,
the dispersion is still quadratic near the top of the hole band.
There is also an ``accidental'' Dirac cone in Fig.
(\ref{fig4}) near the $\Gamma$-point, at a momentum
where the dispersion of the SDW mixed band intersects with that
of the hole band which does not participate  in  the SDW mixing.
This intersection need not to be at the FS  but occurs close to the
FS for the parameters chosen in Fig. (\ref{fig4}).

In Fig. \ref{fig4} we assumed that a particular SDW order
($(0, \pi)$ in that figure) is the same in the whole sample. It is quite possible, though, that
the system has domains with $(\pi,0)$ and $(0,\pi)$ orders. This idea has been recently put forward in Ref. \cite{Yi}.  In Fig. \ref{fig5} we show
the dispersions, the spectral functions, and the
ARPES intensities for a multi-domain sample, assuming equal distributions of the domains with $(\pi,0)$ and ($0,\pi$) orders. Comparing Figs. \ref{fig4} and \ref{fig5}, we clearly see that main effect of averaging over two different SDW orders fis the appearance of an dditional electron dispersion near the $M$-point.
This second electron dispersion is consistent with recent ARPES measurements
of the electronic structure in $BaFe_2As_2$ by Shen and collaborators.~\cite{Yi}

\section{Summary}
\label{sec:6}

To summarize, in this paper we analyzed SDW order in the
itinerant model for Fe-pnictides. We considered a model consisting of
two hole bands centered at $(0,0)$ and two electron bands centered at $(0,\pi)$ and $(\pi,0)$ points in the unfolded BZ (in the folded zone, this corresponds to
two  hole bands centered at $(0,0)$ and two electron bands centered at $(\pi,\pi)$).

We assumed that one hole band is more strongly
 coupled to electron bands and considered first a 3 band model consisting of one hole and two electron bands. We found that SDW order in this model is highly degenerate for the case of a perfect nesting, and if we restrict  with only
the interactions between holes and electrons which
give rise to a SDW instability.
The degeneracy of the SDW order parameter is due to the fact that the energy gain to create SDW order parameters with momentum ${\bf Q}_1 = (0,\pi)$ or ${\bf Q}_2 = (\pi,0)$ is the same.
The order parameter manifold consists of states with a perfect antiferromagnetic order along diagonals but with
non-equal values and arbitrary directions of the OPs at nearest neighbors.

We demonstrated that the degeneracy is broken once we include into consideration  interactions between two electron pockets and ellipticity of these pockets. Both the interactions and the ellipticity select $(0,\pi)$ or $(\pi,0)$  stripe phases with only one ordering  momenta, either  ${\bf Q}_1$ or ${\bf Q}_2$. These stripe-ordered states are in agreement with the experiments on $FeAs$-based pnictides.  The selection of either $(0,\pi)$ or $(\pi,0)$ order leaves one electron FS intact, and the system remains a metal even when SDW order is strong, and the two bands involved in the SDW mixing are fully gapped.

We argued that the selection of stripe states only occurs when the interaction
 between electron pockets is in the charge channel.
We considered, as an example, the model with the interaction in the spin channel and showed that there electron-electron interactions select a different state.

We next added the second hole band and argued that it mixes with the electron band which was left out of the SDW mixing in the 3 band model, once the corresponding coupling exceeds the critical value.  We found, however,
 that this second SDW order necessary gives rise to a non-stripe spin configuration in which both ordering momenta  ${\bf Q}_1$ or ${\bf Q}_2$ are present. The stripe  $(0,\pi)$ or $(\pi,0)$ order is therefore preserved only
if the interaction involving the second hole band
 is below the threshold, and the second SDW order does not develop.

This in turn implies that, if the SDW order is  $(0,\pi)$ or $(\pi,0)$, as
the experiments indicate, only one hole and one electron FSs are involved in the SDW mixing. Other two bands (one hole band and one electron band) are not involved and remain the same as in the normal state. As a result, the system remain a metal for any coupling strength.

We argued that this peculiar requirement is related to the fact that both hole FSs are centered at $(0,0)$. If one of two hole FSs was instead centered at $(\pi,\pi)$, the second SDW order would have the same momentum as the first one
(either  ${\bf Q}_1$ or ${\bf Q}_2$), and the stripe order would survive.
To illustrate this point, we considered a half-filled $t-t--U$ model in which
 the two hole bands are centered at $(0,0)$ and $(\pi,\pi)$.
 In the large $U$ limit,
the model reduces to $J_1-J_2$ model of localized spins.
We argued that, in this model,  the degeneracy of the found state SDW order parameter manifold
 is not broken by either ellipticity or electron-electron interactions, and
 the degenerate manifold includes  $(0,\pi)$ and $(\pi,0)$ states even when
 all four bands are involved in the SDW mixing.
The degeneracy is broken by fluctuations
beyond mean-field, which likely select the stripe order at any $U$.
Because all 4 bands are involved, the stripe-ordered state in the $t-t'-U$ modelit is a metal at small $U$ and an insulator at large $U$.

We analyzed  the electronic structure for parameters relevant to the pnictides, for which SDW order is moderate, and the two FSs involved in the SDW mixing are only partly gapped.  We found  three bands near $k=(0,0)$ and three bands near $k= (\pi,\pi)$ in the folded BZ, and more FS crossings than in the paramagnetic state.  We calculated ARPES intensity and found a number of features
 consistent with the data. In particular, we found  ``Dirac points'' in the dispersion near $(0,0)$ and  an electron band and two hole ``half-bands'' (hole "blades") near $(\pi,\pi)$.

We believe that the good  agreement with
ARPES  experiments is a  strong argument in favor of the itinerant scenario for the ferropnictides. We emphasize that the itinerant scenario does not imply that the system must be in a weak coupling regime. Interactions in the pnictides are
moderately strong and, quite possibly, give rise to some redistribution of the
 spectral weight up to high energies~\cite{si,kotliar,mazin_last}.
The only requirement for the applicability of  our itinerant approach is the existence of a substantial spectral weight  at low-energies, where the system behaves as an interacting Fermi liquid.

\section{Acknowledgements}

We thank E. Abrahams, W. Brenig, S. Borisenko,  P. Brydon,
D. Evtushinsky, A. Kordyuk, J. Knolle, I. Mazin, Q. Si, O. Sushkov, N. Shannon, Z. Tesanovic, C. Timm, M. Vavilov, A. Vorontsov, G. Uhrig, and V. B. Zabolotnyy for useful discussions.
We are particularly thankful to Z. Tesanovic for detecting an error in the early version of the paper.  I.E. acknowledges the support from  the RMES Program (Contract No. N
2.1.1/3199), NSF ICAM Award (Grant DMR-0456669), and thanks  for hospitality UW Madison where this
work was initiated. A.V.C. acknowledges the support from NSF-DMR 0906953,
MPI PKS in Dresden, and Aspen Center for Theoretical Physics, where this work has been completed.


\begin{thebibliography}{99}

\bibitem{kamihara} Y. Kamihara, T. Watanabe, M. Hirano, and H.
Hosono, J. Am. Chem. Soc. {\bf 130} 3296 (2008).
\bibitem{ext_s} I.I. Mazin, D.J. Singh, M.D. Johannes, and M.H. Du, \prl {\bf 101}, 057003 (2008); K. Kuroki, S. Onari, R. Arita, H. Usui, Y. Tanaka, H. Kontani, and H. Aoki,
\prl \textbf{101}, 087004 (2008);
V. Barzykin and L.P. Gorkov, JETP Lett. {\bf 88}, 142 (2008).
\bibitem{maier} T. A. Maier, S. Graser, D. J. Scalapino, and P. J. Hirschfeld,
Phys. Rev. B 79, 224510 (2009); A.V. Chubukov, M.G. Vavilov, and
A. B. Vorontsov, Phys.  Rev. B {\bf 80}, 140515(R) (2009).

\bibitem{kaminski} C. Liu, G.D. Samolyuk, Y. Lee, N. Ni, T. Kondo, A.F. Santander-Syro, S.L. Bud'ko, J.L. McChesney, E. Rotenberg, T. Valla, A. V. Fedorov, P.C. Canfield, B.N. Harmon, A. Kaminski,
 Phys. Rev. Lett. {\bf 101}, 177005 (2008);  D.V. Evtushinsky, D.S. Inosov, V.B. Zabolotnyy, A. Koitzsch, M. Knupfer, B. B\"uchner, M.S. Viazovska, G.L. Sun, V. Hinkov, A.V. Boris, C.T. Lin, B. Keimer, A. Varykhalov, A.A. Kordyuk, and S.V. Borisenko, Phys. Rev. B {\bf 79}, 054517 (2009);  D. Hsieh, Y. Xia, L. Wray, D. Qian, K. Gomes, A. Yazdani, G.F. Chen, J.L. Luo, N.L. Wang, and M.Z. Hasan,
 arXiv:0812.2289 (unpublished);  H. Ding, K. Nakayama, P. Richard, S. Souma, T. Sato, T. Takahashi, M. Neupane, Y.-M. Xu, Z.-H. Pan, A.V. Federov, Z. Wang, X. Dai, Z. Fang, G.F. Chen, J.L. Luo, N.L. Wang,
 arXiv:0812.0534 (unpublished).

\bibitem{coldea}
A.I. Coldea, J.D. Fletcher, A. Carrington, J.G. Analytis, A.F. Bangura, J.-H. Chu, A.S. Erickson, I.R. Fisher, N.E.
Hussey, and R.D. McDonald, Phys. Rev. Lett. {\bf 101}, 216402 (2008).

\bibitem{cruz} Clarina de la Cruz, Q. Huang, J. W. Lynn, J. Li, W. Ratcliff II, J.L. Zarestky, H.A. Mook, G.F. Chen, J.L. Luo, N.L. Wang, and P. Dai,
Nature {\bf453}, 899 (2008).

\bibitem{klauss} H.-H. Klauss, H. Luetkens, R. Klingeler, C. Hess, F.J. Litterst, M. Kraken, M. M. Korshunov, I. Eremin, S.-L. Drechsler, R. Khasanov, A. Amato, J. Hamann-Borrero, N. Leps, A. Kondrat, G. Behr, J. Werner, B. B\"uchner,
Phys. Rev. Lett. {\bf 101}, 077005 (2008).

\bibitem{chandra} P. Chandra, P. Coleman and A.I. Larkin, Phys. Rev. Lett.
{\bf 64}, 88 (1990).

\bibitem{si} Q. Si and E. Abrahams, Phys. Rev. Lett. 101, 076401 (2008);
C. Fang, H. Yao, W.-F. Tsai, J.P. Hu, and S.A. Kivelson, Phys. Rev. B {\bf 77}, 224509 (2008); Cenke Xu, Markus Muller, and Subir Sachdev, Phys. Rev. B 78, 020501(R) (2008); T. Yildirim, Phys. Rev. Lett. {\bf 101}, 057010 (2008);
Goetz S. Uhrig, Michael Holt, Jaan Oitmaa, Oleg P. Sushkov, and
Rajiv R. P. Singh, Phys. Rev. B {\bf 79}, 092416 (2009).

\bibitem{Tesanovic} V. Cvetkovic and Z. Tesanovic, EPL {\bf 85}, 37002 (2009).
See also  V. Stanev, J. Kang, and Z. Tesanovic, Phys. Rev. B 78, 184509 (2008).

\bibitem{lp} V. Barzykin and  L. P. Gorkov, JETP Letters 88, 131 (2008).

\bibitem{Chubukov2008} A.V. Chubukov, D.V. Efremov, and I. Eremin, Phys. Rev. B \textbf{78}, 134512 (2008).
\bibitem{d_h_lee} Fa Wang, Hui Zhai, Ying Ran, Ashvin Vishwanath, and Dung-Hai Lee, Phys. Rev. Lett. {\bf 102}, 047005 (2009).
\bibitem{Korshunov2008} M.M. Korshunov and I. Eremin, Phys. Rev. B
    \textbf{78}, 140509(R) (2008); Europhys. Lett. \textbf{83}, 67003 (2008).
\bibitem{timm} P.M.R. Brydon and C. Timm, Phys. Rev. B {\bf 79}, 180504(R) (2009).
\bibitem{mj} M. D. Johannes and  I. Mazin, arXiv:0904.3857 (unpublished). See also I.I. Mazin and  M.D. Johannes, Nat. Phys. \textbf{5},141
(2009).

\bibitem{honerkamp} C. Platt, C. Honerkamp, and W. Hanke, arXiv:0903.1963(unpublished).

\bibitem{rice}
 M.T. Rice, Physical Review B {\bf 2},  3619  (1970); N. Kulikov and V.V.Tugushev, Sov. Phys. Usp. {\bf 27},  954  (1984),
  [Usp.Fiz.Nauk {\bf 144} 643 (1984)] and references therein. See also
 L.V. Keldysh and Yu.V. Kopaev, Sov. Phys.--Sol. State Phys. \textbf{6}, 2219 (1965); J. De Cloizeaux, J. Phys. Chem. Sol. \textbf{26}, 259 (1965); B.I. Halperin and M.T. Rice, Sol. State Phys. \textbf{21}, 125 (1968).

\bibitem{LDA} D.J. Singh and M.-H. Du, Phys. Rev. Lett. \textbf{100}, 237003 (2008); L. Boeri, O.V. Dolgov, and A.A. Golubov, Phys.
    Rev. Lett. \textbf{101}, 026403 (2008); I.I. Mazin, D.J. Singh, M.D. Johannes, and M.H. Du, Phys. Rev. Lett. \textbf{101}, 057003 (2008).
\bibitem{Coldea2008} A.I. Coldea, J.D. Fletcher, A. Carrington, J.G. Analytis, A.F. Bangura, J.-H. Chu, A.S.Erickson, I.R. Fisher, N.E. Hussey, and R.D. McDonald, Phys. Rev. Lett. \textbf{101}, 216402 (2008).
\bibitem{Lu2008} D.H. Lu, M. Yi, S.-K. Mo, A.S. Erickson, J. Analytis, J.-H. Chu, D.J. Singh, Z. Hussain, T.H. Geballe, I.R. Fisher, and Z.-X. Shen,
    Nature \textbf{455}, 81 (2008).
\bibitem{Liu2008} C. Liu, T. Kondo, M.E. Tillman, R. Gordon, G.D. Samolyuk, Y. Lee, C. Martin, J.L. McChesney, S. Bud'ko, M.A. Tanatar, E. Rotenberg, P.C. Canfield, R. Prozorov, B.N. Harmon, and A.Kaminski,
    arXiv:0806.2147 (unpublished).
\bibitem{Liu2008_2} C. Liu, G.D. Samolyuk, Y. Lee, N. Ni, T. Kondo, A.F. Santander-Syro, S.L. Bud'ko, J.L. McChesney, E. Rotenberg, T. Valla, A.V. Fedorov, P.C. Canfield, B.N. Harmon, and A.Kaminski,
    Phys. Rev. Lett. \textbf{101}, 177005 (2008).
\bibitem{ding_latest} H. Ding, K. Nakayama, P. Richard, S. Souma, T. Sato, T. Takahashi, M. Neupane, Y.-M. Xu,Z.-H. Pan, A.V. Fedorov, Z. Wang, X. Dai, Z. Fang, G.F. Chen, J.L. Luo, and N.L. Wang,
    arXiv:0812.0534 (unpublished).

\bibitem{mazin_last}
S. J. Moon, J. H. Shin, D. Parker, W. S. Choi, I. I. Mazin, Y. S. Lee, J. Y. Kim, N. H. Sung, B. K. Cho, S. H. Khim, J. S. Kim, K. H. Kim, and T. W. Noh,
arXiv:0909.3352.

\bibitem{coldea_nesting} A.I. Coldea, C.M.J. Andrew, J.G. Analytis, R.D. McDonald, A.F. Bangura, J.-H. Chu, I.R. Fisher, and A. Carrington, Phys. Rev. Lett. {\bf 103}, 026404 (2009).

\bibitem{fetese} Y. Xia, D. Qian, L. Wray, D. Hsieh, G.F. Chen, J.L. Luo, N.L. Wang, and M.Z. Hasan,
Phys. Rev. Lett. {\bf 103}, 037002 (2009).

\bibitem{kondo} Takeshi Kondo, R. M. Fernandes, R. Khasanov, Chang Liu, A. D. Palczewski, Ni Ni, M. Shi, A. Bostwick, E. Rotenberg, J. Schmalian, S. L. Bud'ko, P. C. Canfield, A. Kaminski, 	arXiv:0905.0271 (unpublished)

\bibitem{zhou} G. Liu, H. Liu, L. Zhao, W. Zhang, X. Jia, J. Meng, X. Dong, G. F. Chen, G. Wang, Y. Zhou, Y. Zhu, X. Wang, Z. Xu, C. Chen, and X. J. Zhou, arXiv:0904.0677 (unpublished).

\bibitem{dresden_arpes}
V. B. Zabolotnyy, D. S. Inosov, D. V. Evtushinsky, A. Koitzsch, A. A. Kordyuk, G. L. Sun, J. T. Park, D. Haug, V. Hinkov, A. V. Boris, C. T. Lin, M. Knupfer, A. N. Yaresko, B. Buechner, A. Varykhalov, R. Follath, S. V. Borisenko
, Nature 457, 569 (2009).

\bibitem{Yi} M. Yi, D.H. Lu, J.G. Analytis, J.-H. Chu, S.-K. Mo, R.-H. He, M. Hashimoto, R.G. Moore,
I.I. Mazin, D.J. Singh, Z. Hussain, I.R. Fisher, Z.-X. Shen, arXiv:0909.0831 (unpublished).

\bibitem{CT09}  V. Cvetkovic and Z. Tesanovic, unpublished;
Z. Tesanovic, private communication.

\bibitem{barzykin} Victor Barzykin and Lev P. Gor'kov, Phys. Rev. B {\bf 79}, 134510 (2009).

\bibitem{parker} D. Parker, M.G. Vavilov, A.V. Chubukov, I.I. Mazin, Phys. Rev. B {\bf 80}, 100508 (2009).


\bibitem{swz} J. R. Schrieffer, X. G. Wen, and S. C. Zhang, Phys. Rev. B 39, 11663 (1989);
 A. Singh and Z. Tesanovic, Phys. Rev. B 41, 614 (1990);
 A.V. Chubukov and D.M. Frenkel, Phys. Rev. B 46, 11884 (1992).

\bibitem{vishwanath} Y. Ran, F. Wang, H. Zhai, A. Vishwanath, and D.-H. Lee, Phys. Rev. B {\bf 79}, 014505 (2009).

\bibitem{vorontsov} A.B.Vorontsov, M.G.Vavilov, A.V.Chubukov, Phys. Rev. B {\bf 79}, 060508(R) (2009).

\bibitem{brydon} P.M.R. Brydon, and C. Timm, Phys. Rev. B {\bf 80}, 174401 (2009).

\bibitem{tohyama} E. Kaneshita, T. Morinari, T. Tohyama, arXiv:0909.1081 (unpublished).

\bibitem{jose} J. Lorenzana, G. Seibold, C. Ortix, and M. Grilli, Phys. Rev. Lett. {\bf 101}, 186402 (2008).

\bibitem{richard} P. Richard, K. Nakayama, T. Sato, M. Neupane, Y.-M. Xu, J. H. Bowen, G. F. Chen, J. L. Luo, N. L. Wang, H. Ding, T. Takahashi, arXiv:0909.057 (unpublished).

\bibitem{suchitra} N. Harrison, S. E. Sebastian,
arXiv:0910.4199, unpublished.

\bibitem{kotliar}
A. Georges, G. Kotliar, W. Krauth, and M. J. Rozenberg,
Rev. Mod. Phys. 68, 13 (1996).


\end{thebibliography}
\end{document}